\documentclass[10pt,journal,compsoc]{IEEEtran}

\pdfoutput=1

\ifCLASSOPTIONcompsoc
  \usepackage[nocompress]{cite}
\else
  \usepackage{cite}
\fi
\usepackage{textcomp}
\usepackage{stfloats}
\usepackage{url}
\usepackage{verbatim}
\usepackage{graphicx}
\usepackage{cite}
\newcommand\TODO[1][]{{\color{orange}[TODO\ifthenelse{\equal{#1}{}}{}{: #1}]}}
\usepackage{natbib}
\setcitestyle{numbers,square}
\usepackage{amsmath}
\usepackage{amssymb}
\usepackage{mathtools}
\usepackage{amsthm}
\usepackage[utf8]{inputenc} 
\usepackage[T1]{fontenc}    
\usepackage{hyperref}       
\usepackage{url}            
\usepackage{booktabs}       
\usepackage{amsfonts}       
\usepackage{nicefrac}       
\usepackage{microtype}      
\usepackage{xcolor}         
\usepackage{amsmath}
\usepackage{graphicx}
\usepackage{multicol, multirow}
\usepackage{colortbl}
\usepackage[ruled,vlined,linesnumbered]{algorithm2e}
\usepackage{algpseudocode}
\usepackage{subcaption}
\usepackage{ulem}
\usepackage{makecell}
\usepackage{colortbl}

\usepackage{resizegather}

\newcommand{\ie}{\emph{i.e.}}
\newcommand{\eg}{\emph{e.g.}}

\newcommand{\method}{MADE}

\SetCommentSty{mycommentfont}

\newtheorem{problem}{Problem}

\newcommand{\clnloss}[0]{\mathcal{L}_\textrm{cln}}
\newcommand{\natloss}[0]{\mathcal{L}_\textrm{nat}}
\newcommand{\advloss}[0]{\mathcal{L}_\textrm{adv}}
\newcommand{\smhloss}[0]{\mathcal{L}_\textrm{smh}}
\newcommand{\clnV}[0]{\mathcal{V}_\textrm{clean}}
\newcommand{\homoMean}{\mu^\textrm{homo}}
\newcommand{\homoStd}{\sigma^\textrm{homo}}

\begin{document}
%


\title{\method{}: Graph Backdoor Defense \\ with Masked Unlearning}

%
%
%
%

\author{Xiao Lin, Mingjie Li, Yisen Wang
\IEEEcompsocitemizethanks{\IEEEcompsocthanksitem Xiao Lin was at Peking University when the work was completed. Now he is a Ph.D. at IdeaLab, Department of Computer Science, University of Illinois at Urbana-Champaign. Email: xiaol13@illinois.edu. \protect\\
\IEEEcompsocthanksitem Mingjie Li was at Peking University when the work was completed. Now he is a postdoc at CISPA Helmholtz Center for Information Security. Email: lmjat0111@outlook.com.\protect\\
\IEEEcompsocthanksitem Yisen Wang is with State Key Laboratory of General Artificial Intelligence, School of Intelligence Science and Technology, Peking University. Email: yisen.wang@pku.edu.cn. (Yisen Wang is the corresponding author)\protect\\
}
\thanks{Manuscript received April 19, 2005; revised August 26, 2015. 
}
}

%
%

\markboth{Journal of \LaTeX\ Class Files,~Vol.~14, No.~8, August~2015}%
{Shell \MakeLowercase{\textit{et al.}}: Bare Advanced Demo of IEEEtran.cls for IEEE Computer Society Journals}
%



\IEEEtitleabstractindextext{%
\begin{abstract}
Graph Neural Networks (GNNs) have garnered significant attention from researchers due to their outstanding performance in handling graph-related tasks, such as social network analysis, protein design, and so on.
Despite their widespread application, recent research has demonstrated that GNNs are vulnerable to backdoor attacks,
implemented by injecting triggers into the training datasets. Trained on the poisoned data, GNNs will predict target labels when attaching trigger patterns to inputs.
This vulnerability poses significant security risks for GNNs' applications in sensitive domains, such as drug discovery. While there has been extensive research into backdoor defenses for images, strategies to safeguard GNNs against such attacks remain underdeveloped. Furthermore, we point out that conventional backdoor defense methods designed for images cannot work well when directly implemented on graph data. In this paper, we first analyze the key difference between image backdoor and graph backdoor attacks. Then we tackle the graph defense problem by presenting a novel approach called \method, which devises an adversarial mask generation mechanism that selectively preserves clean sub-graphs and further leverages masks on edge weights to eliminate the influence of triggers effectively.
Extensive experiments across various graph classification tasks demonstrate the effectiveness of \method{} in significantly reducing the attack success rate (ASR) while maintaining a high classification accuracy.
\end{abstract}

}

\maketitle

\IEEEdisplaynontitleabstractindextext

%
\IEEEpeerreviewmaketitle

\section{Introduction}

Graph data has recently gained significant attention for their ubiquity, ranging from social networks~\cite{fan2019graph, newman2002random} to protein structures \cite{trinajstic2018chemical}. 
The distinctive feature of graph data lies in their topology structure, which significantly influences final predictions. Notably, a node's class often exhibits strong connections with the labels of its neighbors \cite{zhu2020beyond, wang2022powerful}, and causal relationships may exist between a graph's specific structure and its properties, especially in molecular analysis and other domains. Traditional deep neural networks like convolution neural networks or multilayer perceptrons often fall short in making accurate predictions, particularly when dealing with smaller graph datasets, as they struggle to leverage the rich information embedded in the graph structure for analysis.

To address these challenges, Graph Neural Networks (GNNs) have emerged as a widely adopted learning framework for graph data \cite{kipf2016semi,hamilton2017inductive,li2016gated}. GNNs employ a message-passing mechanism to learn the structural relationships, aggregating information from the local neighborhood of each node. This allows GNNs to capture crucial topological and attributive features inherent in graph data. As a result, GNNs have proven effective across diverse application domains related to graph structures, including drug discovery \citep{lim2019predicting, lin2020kgnn, jiang2021could}, traffic forecasting \citep{jiang2022graph}, 3D object detection \citep{shi2020point}, recommender systems \citep{fan2019graph, wang2019neural}, and webpage ranking \cite{bojchevski2019pagerank, klicpera2018predict}.


However, recent works \cite{xi2021graph,zhang2021backdoor,yang2022transferable} show that GNNs are susceptible to backdoor attacks. Attackers can easily manipulate the model's predictions by inserting specific graph patterns for harmful purposes, known as triggers, into the original graphs. For example, attackers can surreptitiously append trigger subgraphs or features to training graphs, leading GNNs employed in social network analysis or molecular studies to produce predictions that align with the attackers' goals. This vulnerability significantly undermines the reliability of GNNs, especially in high-stakes applications such as protein structure prediction and drug discovery. Therefore, interest in effective methods to mitigate these problems increased sharply these days.


Unfortunately, due to some unique features of graph tasks compared with image and textual tasks, the commonly used backdoor defense methods, like adversarial neuron pruning \cite{wu2021adversarial} (ANP) and Anti-backdoor Learning \cite{li2021anti} (ABL), demonstrate unsatisfied performance on graph data. The first unique feature of the graph task is that its graph structure contains more information compared with image and textual data. For example, in graph classification, different nodes can be connected with an edge due to some specific reasons like social communication, chemical properties, and others, while different components in image or textual data can only be connected by spatial or time series. Such a unique feature gives large flexibilities for attacks and enforces the defending procedure to be more careful like specifying the harness of every edge and node in the graph. The second feature is that graph datasets are much harder to collect and thus usually significantly smaller compared with image and textual datasets, which introduces new difficulties for backdoor defense since the impacts of single backdoor data are much more significant.

\begin{table*}[t]
\caption{Comparisons of current backdoor defense methods on graph backdoors.}
\centering
\resizebox{\linewidth}{!}{\begin{tabular}{c|cccc}
    \toprule
     \textbf{Methods}  & \textbf{Training Set Only}  & \textbf{End-to-End} & \textbf{Poisoned Instance Detection} & \textbf{Poisoned Sub-Graph detection} \\
    \midrule
      Adversarial Neuron Pruning~\cite{wu2021adversarial}     & $\times$ & $\times$ & $\times$ & $\times$ \\
      Finetuning~\cite{sha2022fine} & $\times$ & $\times$ & $\times$ & $\times$
      \\
      Anti-Backdoor Learning~\cite{li2021anti}& $\checkmark$ & $\checkmark$ & $\checkmark$ & $\times$
      \\
      \midrule
      \method{} & $\checkmark$ & $\checkmark$ & $\checkmark$ & $\checkmark$ \\
    \bottomrule
\end{tabular}}

\label{tab:requirements}
\end{table*}

To address the challenge of defending against backdoor attacks on graphs considering these unique features, we propose our approach called \textit{\underline{M}ask Gr\underline{a}ph \underline{De}fense} (\method), illustrated in Figure \ref{fig:pipeline}, which can effectively mitigate backdoor attacks on graphs. Firstly, we construct a module for \textbf{poisoned sub-graph detection}, which can effectively handle the harmful edges and nodes being poisoned instead of only detecting the poisoned instances. 
Secondly, we give up the strategy of utilizing additional clean data to purify the attacked model due to its extremely large workload. Instead, we only aim to obtain a clean model based on a given poisoned training dataset=, which we called \textbf{Training Set Only}.
Thirdly, \method{} is an \textbf{end-to-end} approach that adaptively learns to distinguish between poisoned features and clean features on the poisoned data.
We also compared different defense methods in Table \ref{tab:requirements}. From the table, one can see that our \method{} not only requires fewer additional conditions but also does more accurate trigger sub-graph detection to help GNNs generalize better during training and achieve satisfying performance on both graph and node classification methods. 

In summary, our main contributions can be summarized as follows:

\begin{itemize}
    \item We first discuss the difference between backdoor defenses in graph and image domains and point out that training time defense is the more practical setting. Then we reveal the failure of vanilla training-time defense methods relying on the unsuitable design of their two key components, backdoor isolation and removal.

    \item To tackle the poor performance on backdoor sample isolation, we explore the relationship between backdoor graphs and their homophily behaviors. Then we propose a better backdoor isolation method on graph datasets with significantly better performance. 

    \item We comprehensively study the usefulness of graph topology in backdoor attacks and manage to adopt learnable masks to topologically remove triggers and mitigate their influence instead of using the unlearning paradigm. Combined with the new isolation methods, we propose \method, a training-time defense mechanism without using additional clean datasets.
    
    \item We conduct extensive experiments on four widely used graph datasets for graph and node classifications, demonstrating that \method{} achieves state-of-the-art performance. Our results show that \method{} significantly reduces the attack success rate (ASR) of backdoor attacks while maintaining a high classification accuracy. 
\end{itemize}

\section{Preliminary}
In this section, we first briefly introduce some typical GNN models and basic concepts of backdoor attacks. Then, we further discuss the problems we try to conquer in the following.

\subsection{Graph Neural Networks} \label{sec:gnn_intro}
We first give a detailed definition of the main notation used in this paper, and then introduce some basic GNNs.

\textbf{Notation Convention.} We use bold uppercase and lowercase letters to denote matrices (\eg, $\mathbf{A}$) and vectors (\eg, $\mathbf{v}$), respectively. We also use italic letters for scalars (\eg, $d$), and calligraphic letters for sets (\eg, $\mathcal{N}$). In terms of indexing, the $i$-th row of a matrix is denoted as the corresponding bold lowercase letter with the subscript $i$ (\eg, the $i$-row of $\mathbf{X}$ is $\mathbf{x}_i$).

\textbf{Network Frameworks.} Graph neural networks have demonstrated their expressive power in learning representations of graphs. The key technique for the strong power of GNNs is message passing, which iteratively updates the node features by aggregating information from neighbors for every node in graphs. Specifically, let us have a graph $\mathcal{G} = \{\mathcal{V}, \mathbf{A}, \mathbf{X}\}$ with $\mathcal{V}$, $\mathbf{A}$ and $\mathbf{X}$ representing the set of nodes, the adjacency matrix of the graph $\mathcal{G}$, and the node feature matrix, respectively. Then during $k$ message passing, the update of node features can be expressed as:
    \begin{align} \label{eq:gnn_agg}
        &\mathbf{a}_i^{(k)} = \textbf{AGGREGATE} \left(\{\mathbf{h}_j^{(k)}: v_j \in \mathcal{N}(v_i)\} \right), \\
        &\mathbf{h}_i^{(k+1)} = \textbf{COMBINE} \left(\{\mathbf{h}_i^{(k)}, \mathbf{a}_i^{(k)}\} \right),
    \end{align}
where $\textbf{AGGREGATE}$  and $\mathbf{COMBINE}$ represent the aggregation function and the transform operation, $\mathbf{h}_i^{(k+1)}$ is the feature of the node $v_i$ after $k$ message passing ($\mathbf{h}_i^{(1)} = \mathbf{x}_i$ for initialization), $\mathbf{a}_i^{(k)}$ is the intermediate variables of the node $v_i$, and $\mathcal{N}(\cdot)$ represents the set of the $1$-hop neighbors. For example,  Then by iteratively aggregating graph information, GNNs can effectively process the topological information contained in graph datasets and achieve satisfying results on graph tasks. As one can see, the key architecture of GNNs is different from neural networks for image domains like convolutional networks \cite{vgg, resnet} and transformers \cite{vaswani2017attention}. Therefore, the properties for backdoor-attacked samples are also different from former models.

\subsection{Backdoor Attacks}
In this subsection, we introduce the backdoor attack and its variation in the graph domain.

\textbf{Backdoor Attacks} \cite{tran2018spectral, chen2017targeted, gu2017badnets, liu2020reflection}. Given a dataset $\mathcal{D} = \{\mathcal{X}, \mathcal{Y}\}$ with $\mathcal{X}$ and $\mathcal{Y}$ representing the set of samples and corresponding labels, respectively, attackers generate some special and commonly invisible patterns, which are called triggers. Then attackers add those patterns (triggers) into a small portion of samples in the dataset $\mathcal{D}$. After training on the poisoned dataset, models cannot give the correct predictions if the inputs contain triggers while still performing normally when facing clean inputs, \ie, the inputs without triggers.

\textbf{Backdoor Defense} \cite{wang2019neural, wang2022rethinking, wu2021adversarial, guan2022few, li2021anti}. Existing works fall under the categories of either detecting or erasing methods. Detecting methods aim to find out whether the given dataset or the given model is poisoned. These methods typically come with a relatively high detection accuracy. Erasing methods take a step further, and aim to erase the negative impacts of triggers on models. Normally, the lower the Attack Success Rate (ASR) is after erasing, the more effective the erasing method is. In this paper, we focus on the erasing methods and explore an effective way to remove triggers on graphs.

\textbf{Backdoor on Graphs \cite{zhang2021backdoor, yang2022transferable, dai2023unnoticeable}.} Compared with the backdoor attack designed for other types of data, such as images \cite{gu2017badnets, chen2017targeted}, and texts \cite{kwon2021textual, qi2021mind, dai2019backdoor}, the backdoor attack on graphs has its uniqueness. Specifically, the topology structures of graphs are complex and irregular, which implies that topology structures of graphs contain much richer information than images or texts. Thereby, unlike only adding triggers on feature-level triggers $\mathbf{X}_{tri}$ like image, graph backdoors also modify their adjacency matrices $\mathbf{A}$ with trigger subgraphs denoted as $\mathbf{A}_{tri}$. Then the poison data can be formulated as $\left(\mathbf{X}+\mathbf{X}_{tri}, \mathbf{A}\circ\mathbf{A}_{tri}, y_{tar}\right)$, where $\circ$ is the attachment operation. Training on the poison dataset, GNNs will predict the label of any graph as $y_{tar}$ when attaching $\mathbf{A}_{tri}$ and $\mathbf{X}_{tri}$ to the graph's adjacency matrix and node features. 
To sum up, the problem of backdoor attacks on graphs can be defined as follows:

To sum up, the problem of backdoor attacks on graphs can be defined as follows:

\begin{problem}
    Backdoor attack on graph classification tasks
\end{problem}

\noindent \textbf{Given}: (1) a clean dataset $\mathcal{D}$; (2) the target label $y_{tar}$; (3) the injection rate $\alpha$ \footnote{The injection rate $\alpha$ is typically small, usually less than $10\%$.}. 

\noindent \textbf{Attackers' Capability}: The capability of attackers can be summarized as follows:
\begin{enumerate}
\item They can generate an invisible feature trigger $\mathbf{X}_{tri}$ and trigger subgraphs $\mathbf{A}_{tri}$; 
\item They can poison $\alpha$ graphs among the training set by attaching the trigger $\left(\mathbf{X}_{tri}, \mathbf{A}_{tri}\right)$ and flipping their labels to the target label $y_{tar}$.

\end{enumerate}

\noindent \textbf{Attacker's Goal}: After training on a poisoned dataset, for any clean graph $\left(\mathbf{X}, \mathbf{A}, y \right)$, the attack model $f(\cdot)$ will predict it as the target label when the trigger is attached, \ie, 
\begin{equation*}
f(\mathbf{X} \circ \mathbf{X}_{tri}, \mathbf{A} \circ \mathbf{A}_{tri}) = y_{tar}.
\end{equation*} 
Otherwise, predicting the input as the natural label, \ie,
\begin{equation*}
f(\mathbf{X}, \mathbf{A}) = y.
\end{equation*}

\subsection{Problem Definition}
Based on the definition of Problem 1, our objective is to create a robust defense mechanism ensuring that the model trained under the setting of Problem 1 remains insensitive to triggers. Hence, we define the problem of backdoor defense on graphs as follows:

\begin{problem}
    Backdoor defense on graph classification tasks.
\end{problem}

\noindent \textbf{Given}: A poisoned dataset $\mathcal{D}_{poison}$ is given which has been attached by a user-invisible trigger $\left(\mathbf{X}_{tri}, \mathbf{A}_{tri}\right)$ with the target label $y_{tar}$

\noindent \textbf{Capability of Defenders}: We can only control the training process.

\noindent \textbf{Goal}: We aim to ensure the GNNs model free from data poisoning attacks. After training on a poisoned dataset, get a clean model $f(\cdot)$ free from the backdoor threat, which means the model performs well on clean graphs $\left( \mathbf{X}, \mathbf{A}, y\right)$, \ie, 
\begin{equation*}
f(\mathbf{X}, \mathbf{A}) = y.
\end{equation*}
In the meanwhile, the model cannot be manipulated by triggers, \ie, 
\begin{equation*}
(\mathbf{X} \circ \mathbf{X}_{tri}, \mathbf{A} \circ \mathbf{A}_{tri}) \neq y_{tar}.
\end{equation*}

\subsection{Difference between Image and Graph Backdoor Defense} 
\begin{table}[htbp]
    \centering
    \caption{The statistics of image and graph datasets. \textbf{Sample Number} denotes the amount of the training data, \textbf{Nodes $\backslash$ Pixels} denotes the pixel number of each image sample or node number of each graph, and the \textbf{topology} here means the structure of a node or pixel's ego graph. }
    \resizebox{0.45\textwidth}{!}{
    \begin{tabular}{c|c| ccc}
        \toprule
        \multicolumn{2}{c|}{Types} & \textbf{Sample Number} & \textbf{Nodes $\backslash$ Pixels} & \textbf{Topology}  \\
        \midrule
          \multirow{2}{*}{Images} & CIFAR-$10$ & 60,000 & 3072 & Grid\\
          & ImageNet & 1,281,167  & 544,509 & Grid\\
        \midrule
          \multirow{3}{*}{Graph}& AIDS & 2000 & 15.69 & Graph Dependent\\
          & PROTEINS & 1113 & 39.06 & Graph Dependent\\
          & ENZYMES & 600 & 32.63 & Graph Dependent\\
        \bottomrule
    \end{tabular}
    }
    \label{tab: comp}
\end{table}

Apart from the structural difference between graphs and images, two key factors make the effective protection of graph models from data poisoning even harder.

Firstly, as illustrated in Table \ref{tab: comp}, a critical difference between image and graph datasets is the number of samples. Image datasets generally have sample numbers and feature sizes that are magnitudes larger than those in graph datasets. This abundance allows backdoor defense mechanisms greater flexibility to remove a relatively higher number of potentially poisoned samples. In contrast, graph datasets often have limited training samples, and further removing a portion of these samples could significantly degrade model performance. This scarcity of training data in graph tasks makes backdoor defense more challenging, necessitating highly accurate detection of poisoned samples.



Secondly, unlike backdoor attacks for images that modify features, graph-based backdoor attacks can rewire the edges between nodes for attack. Therefore, even if all node features are clean, an attacker can still successfully attack the model with the modified graph's topology. Consequently, existing backdoor detection methods may be ineffective for graphs, as they primarily focus on feature manipulation and lack designs on the graph structure.


Foretunately, graph structure can also provide us with more information for detection, as we can use many graph-related attributes for the defense like homophily metric, and graph attention. 

\section{\method{} Algorithm} \label{method}
Since our training-time defense setting is the same as ABL's scenario \cite{li2021anti}, we first 
analyze ABL's limitations on graphs from the data and training perspective. Based on that, we propose our \method{} with the help of unique graph properties.

\subsection{Data Isolation}\label{sec:data_isolation}

\begin{figure}[htbp]
    \centering
    \begin{minipage}[c]{0.23\textwidth}
    \centering
    \includegraphics[width=\textwidth]{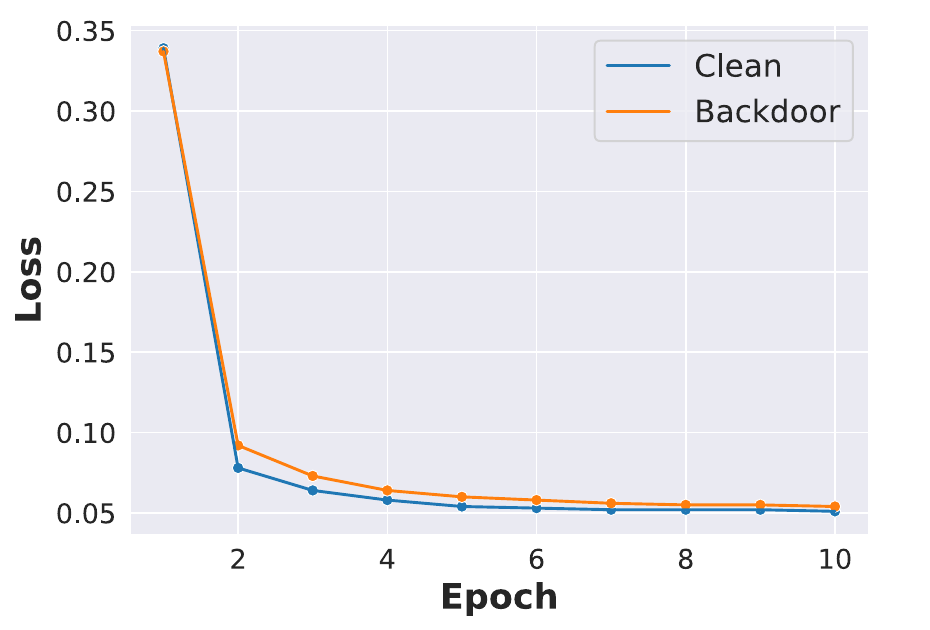}
    \subcaption{GCN}
    \end{minipage}
    \begin{minipage}[c]{0.23\textwidth}
    \centering
    \includegraphics[width=0.98\textwidth]{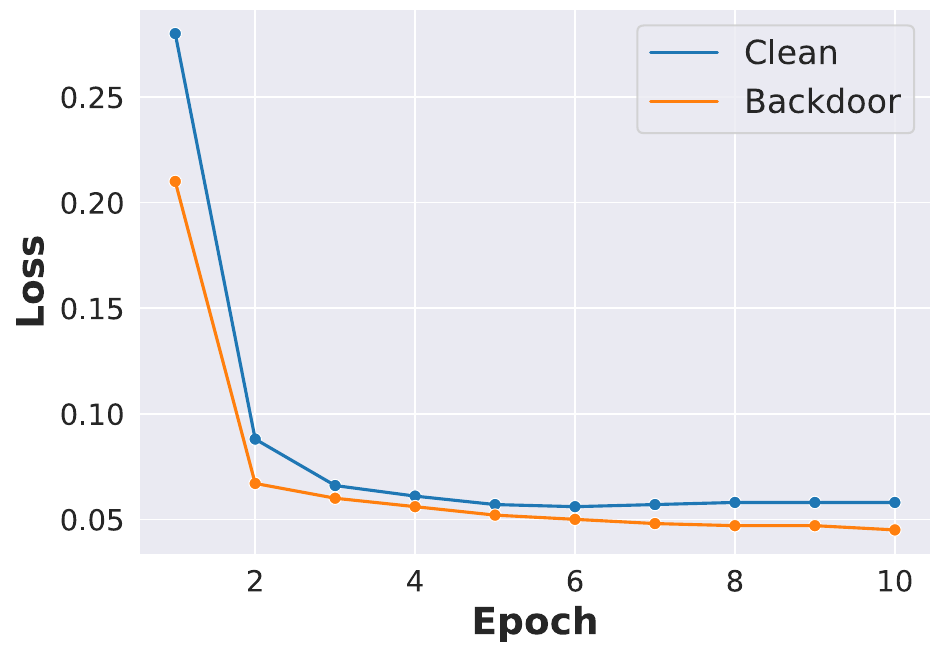}
    \subcaption{GAT}
    \end{minipage}
    \caption{Training curve of PROETINS' backdoor and natural subset on different GNNs for the first $10$ epochs.}
    \label{fig:loss}
\end{figure}

\begin{table}[htbp] 
    \centering
    \caption{Proportion of backdoor sample among enriched backdoor subsets. The higher the proportion, the more effective the data isolation is.}
    \resizebox{\linewidth}{!}{
    \begin{tabular}{c| ccccc}
        \toprule
         & \textbf{AIDS} & \textbf{PROTEINS} & \textbf{PROTEINS\_full} & \textbf{ENZYMES} \\
        \midrule
          \textbf{ABL} & 34.00\% & 89.97\% & 10.11\% & 92.92\% \\
          \textbf{\method{}} & $\mathbf{82.00\%}$ & $\mathbf{94.38\%}$ & $\mathbf{92.1\%}$ & $\mathbf{93.75\%}$ \\
        \bottomrule
    \end{tabular}
    }
    \label{tab:enrich_rate}
\end{table}

\par\noindent

To understand ABL's limitations on graphs, we evaluate the data isolation part in ABL when applying it to graph neural networks. Unfortunately, we find that their isolation accuracy varies significantly across different datasets, as shown in Table~\ref{tab:enrich_rate}. From the above results, one can see that the original ABL's isolation method cannot successfully select suspicious backdoor samples and create an enriched poisoned dataset. Therefore, the performance of ABL is not satisfactory on some graph backdoor datasets. Next, we explore why former isolation methods cannot work and propose effective isolation methods for graphs.

To be specific, we analyze the difference of backdoor samples on the image and graph domains from the training perspective, as shown in Figure \ref{fig:loss}. From the figure, one can see that, unlike image backdoor samples, the difference between the convergence rate for backdoor graphs (orange) and clean graphs (blue) is not significant. 
Therefore, solely relying on loss scores as isolation metrics, as done by ABL, is inadequate in the graph domain and leads to the failure of ABL's application on graph datasets. Therefore, we need to explore new metrics for the backdoor sample's isolation.


\begin{table}[t]
    \centering
    \caption{The average (standard variation) homophily of backdoor graphs and clean graphs.}
    \resizebox{0.45\textwidth}{!}{\begin{tabular}{c| cccc}
        \toprule
         & \textbf{AIDS} & \textbf{PROTEINS} & \textbf{PROTEINS\_full} & \textbf{ENZYMES} \\
        \midrule
          \textbf{Backdoor} & $0.85\pm0.06$ & $0.9\pm0.05$ & $0.81\pm0.08$ & $0.83\pm0.05$ \\
          \textbf{Clean} & $0.98\pm0.02$ & $0.99\pm0.01$ & $0.99\pm0.01$  & $0.99\pm0.02$ \\
        \bottomrule
    \end{tabular}
    }
    \label{tab:homo}
\end{table}


Recently, some researchers~\cite{zhu2022does} have proved that some graph adversarial attacks may inevitably affect the graph's homophily, a widely used metric for evaluating the general relationships between the neighboring nodes. For the graph $\mathcal{G}$, the graph's homophily score $\textrm{homo} (\mathcal{G})$ \cite{li2022finding, zhu2020beyond} can be formulated as:

\begin{equation}\label{eq:graph_homo}
    \textrm{homo}(\mathcal{G}) = \frac{1}{\vert \mathcal{E} \vert} \sum_{(i,j)\in\mathcal{E}} \textrm{sim}(\mathbf{x}_i , \mathbf{x}_k),
\end{equation}
where $\mathcal{E}$ denotes the edge set for the graph $\mathcal{G}$, and $\textrm{sim}(\cdot, \cdot)$ represents the cosine similarity. 

Inspired by the strong connections between the attacked graphs and their homophily changes, we explore the homophily behaviors of clean graphs and backdoor graphs generated by GTA~\cite{xi2021graph}, a typical method of graph backdoor attacks. 
The results presented in Table~\ref{tab:homo} demonstrate a notable difference in homophily scores between backdoor and clean graphs, which can be leveraged to distinguish backdoor graphs from clean ones. Intuitively, since backdoor graphs constitute only a small fraction of the training set, the overall distribution of homophily scores is predominantly influenced by the clean graphs. Consequently, the homophily scores of backdoor graphs exhibit a significant deviation from the average homophily score of the entire training set. Therefore, we can reliably identify the potentially attacked graphs as those whose homophily scores fall within the following deviant range:
\begin{equation}
\begin{aligned}
(0, \homoMean - \homoStd) 
 \cup (\homoMean+\homoStd,1),
\end{aligned}
\end{equation}
where $\homoMean$ and $\homoStd$ represents the mean and standard variance of the homophily scores for all the graphs within the training set, respectively.
As a result, we acquire a subset of graphs with a high possibility of containing triggers, denoted as $\mathcal{D}_{h}$.

However, there are usually not enough outlier graphs to form the detected backdoor subset through the above methods. 
Therefore, additional techniques are required to isolate the remaining graphs. Considering that the sign operator in ABL's warmup loss can impede the convergence of graph models, we opt to discard the sign function in ABL. Instead, we employ the original Cross Entropy loss to train and select the remaining graphs. In this context, a lower classification loss typically signifies a higher likelihood of being backdoor-attacked.
Specifically, following an initial warm-up training phase, we compute the per-sample loss function and select the top samples with the lowest loss as the enriched harmful sample set, denoted as $\mathcal{D}_l$. Combined with $\mathcal{D}_{h}$, we can finally get the enriched backdoor subset denoted as $\mathcal{D}_{bad} = \mathcal{D}_{h} \cup \mathcal{D}_{l}$ with $\alpha_1$ isolation rate. Moreover, we select $\alpha_2$ samples with the highest loss as the enriched clean subset, denoted as $\mathcal{D}_{clean}$. 
To assess the efficacy of our proposed data isolation, we conduct experiments using the GTA backdoor attack with an injection rate of $10.0\%$ on four distinct datasets. we evaluate the proportion of truly injected trigger samples in the enriched backdoor subsets. The experimental results \footnote{In Table 3, the proportion of backdoor samples represents both precision and recall. This equivalence arises from the numerical identity of precision and recall, as the injection and isolation rates are set the same in Table 3. Detailed results about different isolation and injection rates are in Appendix \ref{sec:data_iso_diff_rate_appx} and \ref{sec:data_iso_indiv_method_appx}.}, as shown in Table \ref{tab:enrich_rate}, reveal that our proposed method consistently achieves a relatively high proportion of truly injected trigger graphs, effectively identifying the backdoor samples, especially on PROTEINS\_full and AIDS datasets.

\subsection{Forward Propagation with Masks} \label{sec:forward_prop_mask}
After data isolation, we aim to leverage the enriched clean subset $\mathcal{D}_\textrm{clean}$ to ensure the model's natural classification ability while utilizing the enriched backdoor subset $\mathcal{D}_\textrm{bad}$ to remove the influences of malicious triggers. For graph classification tasks, as attackers usually modify a small portion of nodes within the poisoned graphs to ensure the stealthiness, the isolated poisoned graphs also contain useful structures. Thus, we propose to generate masks for selectively removing the trigger subgraphs while preserving the model's natural performance.

\subsubsection{Masked Aggregation}
\label{sec:graph-mask-agg}

\begin{figure}[htbp]
    \centering
    \begin{minipage}[c]{0.4\textwidth}
    \centering
    \includegraphics[width=\textwidth]{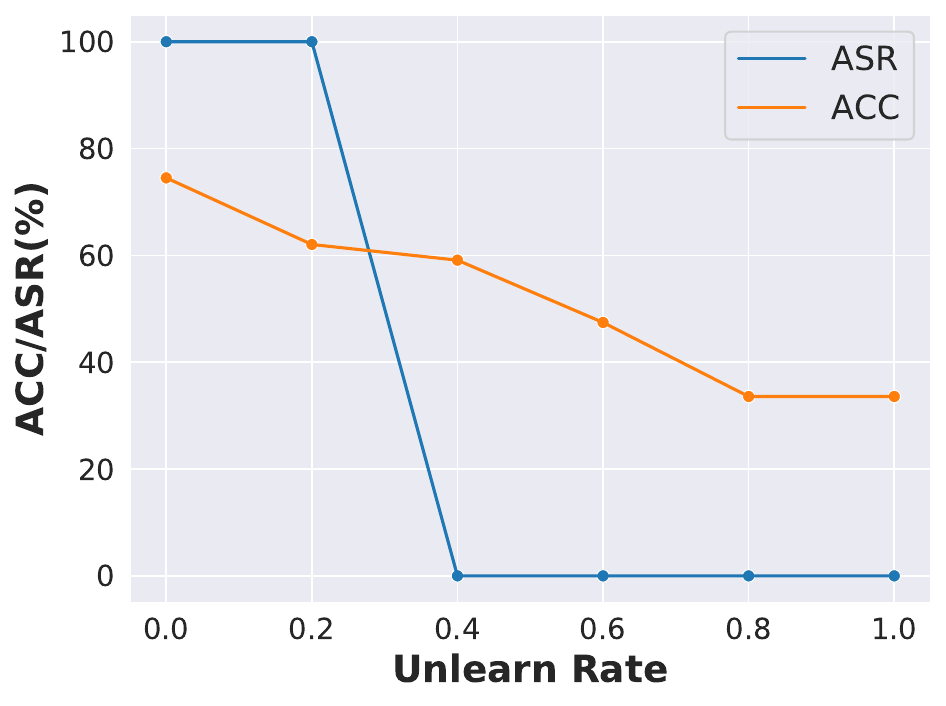}
    \end{minipage}
    \caption{Unlearning performance w.r.t. different unlearn rates on PROTEINS.  The unlearn rates denote the proportion of backdoor samples in the poisoned datasets for unlearning. The orange curve indicates the accuracy (ACC), while the blue curve represents the attack success rate (ASR).}
    \label{fig:unlearn_mask}
\end{figure}

\begin{figure*}[htbp]
    \centering
    \includegraphics[width=\textwidth]{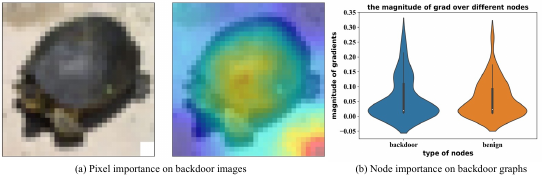}
    \caption{Differences in the response of models attacked by image backdoor and graph backdoor attacks. In Figure 3(a), the left sub-figure displays a poisoned image with a trigger injected in the lower right corner, while the right sub-figure shows a heatmap generated from one attacked model on this poisoned image.  Figure 3(b) showcases the gradients from an attacked GNN across different types of nodes.}
    \label{fig:diff_converge_rate}
\end{figure*}
A straightforward approach to removing backdoor triggers is to mask the entire graphs containing triggers and then unlearn all the masked graphs like ABL \cite{li2021anti} does.
However, as shown in Figure \ref{fig:unlearn_mask}, directly unlearning entire graphs results in a substantial decrease (over $15\%$) in classification accuracy when the influence of triggers is entirely removed, i.e., when the unlearn rate exceeds $0.4$. 
Due to this reason, we are going to separate the backdoor unlearning and natural training procedures into different sub-modules. Firstly, we try to propose a new module to directly purify the input graphs from poisoned ones via a mask generation module. Since the mask generation module and original GNNs are trained separately, the original training process of GNNs will not be modified too much and lead to better performance compared with the original unlearning procedure.  

To purify the backdoor samples, one widely used way in the image domain is to calculate the masks through gradient scores~\cite{selvaraju2017grad} as the model tends to exhibit more straightforward and localized attention to the trigger patterns embedded in the image domain. As shown in Figure \ref{fig:diff_converge_rate}(a), the model's gradient heatmap mainly focuses on the patterns, \ie, the right bottom corner. Unfortunately, such a finding does not apply to the graph domain. As shown in Figure \ref{fig:diff_converge_rate}(b), the distribution of gradient magnitudes for both malicious (backdoor) and clean (benign) nodes across all poisoned graphs are almost the same. The y-axis represents the magnitude values of the gradients, while the x-axis distinguishes between malicious and clean nodes. We can observe that there is a slightly higher proportion of nodes with larger gradient magnitudes in the malicious nodes compared to the clean nodes. However, the overall gradient magnitude distributions for both types of nodes appear relatively similar, without any significant visual difference in Figure \ref{fig:diff_converge_rate} (a). This similarity may be attributed to the more rapid information aggregation facilitated by graph structures compared to image convolutions. Consequently, using gradient scores to generate masks is ineffective in the graph domain.
\begin{figure}[htbp]
    \centering
    \begin{minipage}[c]{0.23\textwidth}
    \centering
    \includegraphics[width=\textwidth]{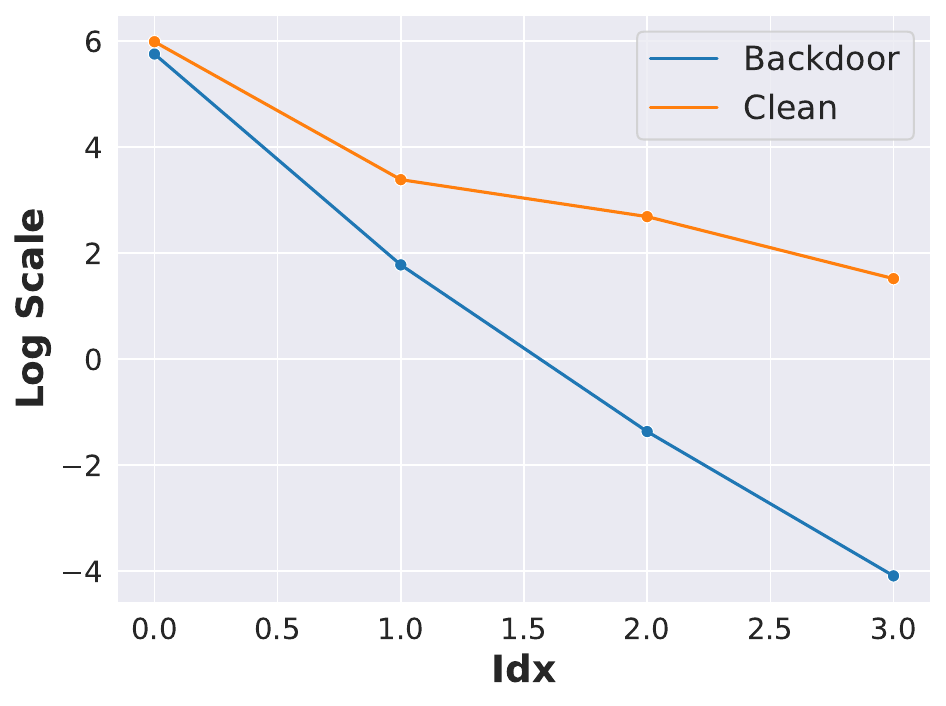}
    \subcaption{PROTEINS.}
    \end{minipage}
    \begin{minipage}[c]{0.23\textwidth}
    \centering
    \includegraphics[width=\textwidth]{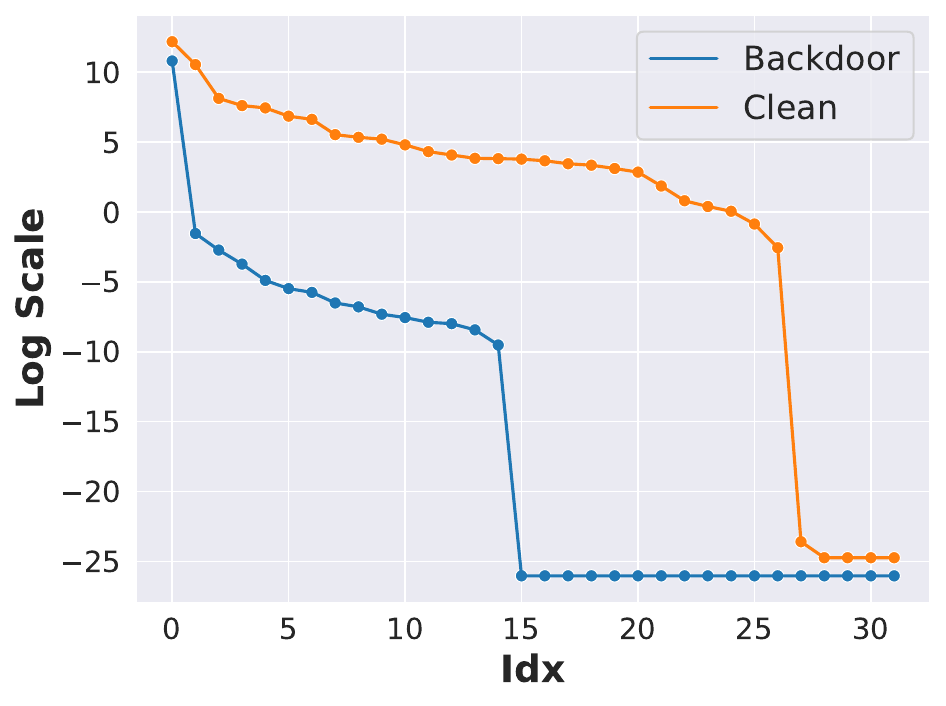}
    \subcaption{PROTEINS\_full.}
    \end{minipage}
    \caption{The spectrum for clean nodes and backdoor nodes on different datasets. Singular values of node feature matrices are computed and sorted in descending order, with the x-axis showing the order (starting from 0) and the y-axis showing log-scaled singular values.}
    \label{fig:spectrum_nodes}
\end{figure}

Fortunately, we manage to find the difference between backdoor samples and natural samples from the spectral perspective, inspired by the former works \cite{wu2022debiasing,tran2018spectral}. Firstly, we draw the features' spectrum of clean nodes and backdoor nodes. Specifically, we stacked the node features of clean and poisoned graphs, respectively, and then computed the singular values of the stacked node features. The sorted singular values are displayed in Figure \ref{fig:spectrum_nodes}. From the figure, it is obvious that the singular values for backdoor nodes decrease rapidly while the singular values for clean nodes distribute more uniformly. The above results indicate that we can effectively separate backdoor subgraphs and natural subgraphs via a linear layer especially if the projection matrix can project node features into a subspace spanned by singular vectors corresponding to small singular values. Then we can generate masks to purify the original graph samples. Thereby, we introduce a learnable projection head that projects node features onto the subspace w.r.t. small singlur values:
\begin{equation}
    p_i^{(k)} = \textbf{HEAD} \left( \mathbf{h}_i^{(k)} \right),
\end{equation}
where $\textbf{HEAD}$ denotes the learnable projection head. From our analysis, the above projection head can be used as a soft indicator of clean nodes or trigger nodes as the projection of the trigger nodes will be close to zero if its weight relates to the eigenvectors of the tail eigenvalues. We leave its training strategy in the following section. However, the scale of projections output may also hinder the node's classification. For instance, a trivial solution to remove backdoors involves setting all nodes' projections to be small during training. While this approach effectively eliminates backdoor nodes, it also significantly diminishes the model's learning ability on clean nodes, thereby degrading overall classification performance. To solve the above problem, we integrate the graph node's neighbors, and hence we define the natural score as below: 
\begin{equation}
    score^{(k)}(v_i) = \textrm{AVG} \left( \textrm{sim}(p_i^{(k)}, p_j^{(k)}) : v_j \in \mathcal{N}(v_i) \right),
\end{equation}
\noindent where $\textrm{AVG}$ denote an average function and $sim(p_i^{(k)}, p_j^{(k)})$ represents the cosine similarity between the node features $\mathbf{h}_i^{(k)}$ and $\mathbf{h}_j^{(k)}$. The neighborhood information can help us judge whether the node is a trigger no matter whether the scaling of $p$ is small. For example, a natural node with a small projection $p$ can still get a higher natural score since all their neighbors are similar and their cosine similarity scores are higher. Conversely, a trigger node's natural score will still be low since they are dissimilar to their neighbor as shown in Table \ref{tab:homo}.  Thus, the natural score can successfully classify whether a node is a natural one or a trigger one.

Furthermore, we notice that backdoor triggers typically occupy only a minor fraction of the overall graph in the pursuit of semantic smoothness and stealthiness. In light of this, we plan to reset edge weights as $1$ to the nodes that are more likely to be clean. By doing so, we can easily ensure the consistency between the original graph topology and the masked one to stabilize the training process, even in cases where certain mask values are relatively diminutive. Specifically, we calculate the natural score $score^{(k)}(v_i)$ for every node $v_i$, and select the top $\beta$ nodes with the highest natural scores, which is represented as $\clnV$. For any node in  $\clnV$, we assign its mask value of 
 $1$.

After classifying the nodes, we extend the node's natural score to edges for the mask generation of the rest nodes. To be specific, we use soft masks to remove the edges connected with trigger nodes and also some outlier edges, whose end nodes are dissimilar to start nodes, as trigger edges can also influence the graph's homophily as demonstrated in Table \ref{tab:homo}. With all the above methods combined, the natural scores for edges (edge masks) are defined below:
\begin{equation} \label{eq:mask_express}
    \mathbf{m}^{(k)}_{i, j} = \left\{
    \begin{aligned}
        & A_{ij},\qquad v_i \in \clnV \  and \  v_j \in \clnV\\
        & sim(p_i^{(k)}, p_j^{(k)})*A_{ij}, \qquad else,
    \end{aligned}
    \right.
\end{equation}
where $\mathbf{m}_{i,j}^{(k)}$ represents the masked adjacency score between node $v_i$ and $v_j$ after $k$ message passing. 
After obtaining $\mathbf{m}$, we treat the graph as a weighted graph for message passing. Mathematically, the message passing is modified as follows:
\begin{align} \label{eq:weight_gnn_agg}
    &\mathbf{a}_i^{(k)} = \textbf{AGGREGATE} \left(\{\mathbf{m}_{i, j}^{(k)} \cdot \mathbf{h}_j^{(k)}: v_j \in \mathcal{N}(v_i)\} \right) \\
    &\mathbf{h}_i^{(k+1)} = \textbf{COMBINE} \left(\{\mathbf{m}_{i, i}^{(k)} \cdot \mathbf{h}_i^{(k)}, \mathbf{a}_i^{(k)}\} \right)
\end{align}



\subsubsection{Loss Definition.} \label{sec:loss_define}
Once we have obtained the mask for the graph, we proceed to update the gradients based on whether the graph belongs to the
enriched poisoned dataset $\mathcal{D}_{bad}$ or the enriched clean dataset $\mathcal{D}_{clean}$.

In the case where the graph $\mathcal{G} = \{ \mathbf{A}, \mathbf{X}\}$ belongs to enriched backdoor subset $\mathcal{D}_{bad}$, we employ an adversarial loss function $\advloss$ to deteriorate the model's performance on the backdoor samples, effectively unlearning triggers. Mathematically, this process can be represented as follows:
\begin{equation}\label{eq:adv_loss}
    \advloss := \textrm{softmax} \left( f \left(\mathbf{A} , \mathbf{X}, \mathbf{m}\right) \right) \left[ y\right]
\end{equation}
where $f(\cdot)$ represents the model to be trained, $\mathbf{m}$ denotes the graph mask mentioned above in Section \ref{sec:forward_prop_mask}, and $y$ is the label. In this case, $y$ is highly likely to be the target label. Therefore, the intuition of Eq. \eqref{eq:adv_loss} is to defend backdoor attack by reducing the probability of successfully predicting backdoor samples.

For the graph $\mathcal{G} = \{ \mathbf{A}, \mathbf{X}\}$ belonging to the enriched clean dataset $\mathcal{D}_{clean}$, we utilize the natural loss $\natloss$ to ensure the natural classification ability of the model on clean graphs. Mathematically, the natural loss could be expressed as:
\begin{equation}\label{eq:natural_loss}
    \natloss := \textrm{CE} \left( \textbf{GNN} \left(\mathbf{A}, \mathbf{X}, \mathbf{m}\right), y \right).
\end{equation}
where $\textrm{CE}$ represents the cross-entropy function, a widely used classification loss.
Besides, to guarantee that the performance of our model on clean samples remains unaltered when masks are added, a supplementary smooth loss $\smhloss$ is incorporated into our training process. This loss serves to harmonize the output of clean samples when processed both with and without masks:
\begin{equation}
 \smhloss :=\left\|f \left(\mathbf{A}, \mathbf{X}, \mathbf{m}\right) -f \left(\mathbf{A}, \mathbf{X} \right)\right\|_2.
\end{equation}
Finally, combining the clean loss $\clnloss$ and the smooth loss $\smhloss$, we arrive at the clean loss $\clnloss$ tailored for clean graphs:
\begin{equation} \label{eq:cln_loss}
    \begin{aligned}
        \clnloss &:= \natloss + \lambda \smhloss 
    \end{aligned}
\end{equation}
where $\lambda$ is a hyperparameter.

\subsection{Method Summarization}

\begin{figure*}[htbp]
    \centering
    \resizebox{\linewidth}{!}{
    \includegraphics{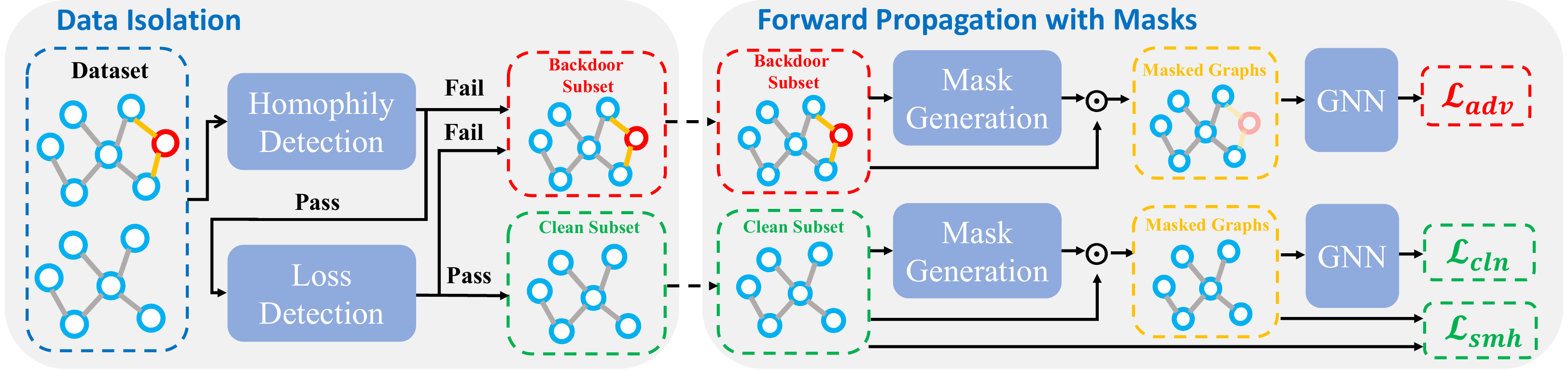}
    }
    \caption{Illustrations of proposed \method. \textcolor{red}{Red circles} / \textcolor{cyan}{blue circles} represent the backdoor nodes / clean nodes, while \textcolor{orange}{orange links} / \textcolor{gray}{gray links} represent backdoor edges / clean edges, respectively. During data isolation, \method{} leverages the discrepancies between backdoor and clean samples in homophily and classification loss to create enriched backdoor subsets and clean subsets. During the forward propagation, \method{} introduces a graph mask learning schema. It employs adversarial loss $Loss_{adv}$ on the backdoor subset to help GNNs forget trigger patterns, while simultaneously utilizing classification loss $Loss_{clean}$ and smoothing loss $Loss_{smh}$ on the clean subset to maintain high classification ability.}
    \label{fig:pipeline}
\end{figure*}

We summarize the above procedures in Figure \ref{fig:pipeline}. From the figure, one can see that training with our method can be roughly divided into two stages: ``Data Isolation'' and ``Forward Propagation with Masks''. Firstly, we try to locate the poisoned subset in the training sets with our homophily and loss detection method. Instead of only finding the poisoned instances like the poisoned graphs or nodes. We try to use a learnable mask generator to remove the poisoned sub-graphs in the second stage and then train the GNN with the purified graphs. After training, our mask generator can also mask the trigger subgraphs of the input graphs and further mitigate the trigger's influence on GNN's predictions. 

Given that \method{} offers instance-level detection (data isolation), it can be seamlessly extended from graph classification tasks to node classification tasks. Please note that, for node classification, the instance-level detection of \method{} can already accurately identify potentially attacked nodes, thus constructing precise masks to filter out triggers. Therefore, for node classification, \method{} can effectively defend backdoor attacks, even without the adaptive mask generation process described in Section \ref{sec:graph-mask-agg}.
Detailed description of \method{} on node classification is provided in Appendix \ref{appdx:node-class-made}.  Algorithm \ref{algo:compre_inf_process} summarizes the inference phase of \method{} for both graph classification and node classification tasks, while Algorithm \ref{algo:training} summarizes the training phase.

\begin{algorithm}[htbp] 
    
    \SetKwInOut{Input}{Input}
	\SetKwInOut{Output}{Output}
    \Input{A graph $\mathcal{G}$ with its adjacent matrix $\mathbf{A}$ and node feature $\mathbf{X}$ with $\mathbf{x}_i$ denotes the node feature for node $i$, a $K$-layer GNN with weight $\mathbf{W}$ trained with our \method{}.}
    \Output{Model's final output $\mathbf{Y}$}
         Initialize $\mathbf{H}_0 = \mathbf{X}$\;
        
        \For{$k$ in [$1$,...,$K$]}{
        \uIf{Doing graph classification}{
        Get layer $k$'s mask $\mathbf{m}^{(k)}$ based on Eq. \eqref{eq:mask_express}\;
        }
        \uElseIf{Doing node classification} {
            Get layer $k$'s mask $\mathbf{m}^{(k)}$ based on Eq. \eqref{eq:mask_express_node}\;
        }
        Get the aggregated features $\mathbf{A}^{(k)}$ based on Eq. \eqref{eq:weight_gnn_agg}\;
        Get the projected embedding $\mathbf{H}^{(k)} = \mathbf{A}^{(k)} \mathbf{W}^{(k)}$\;
    }
    Get final output $\mathbf{H} = \mathbf{H}^{(K)}$.
    \caption{GNNs inference with our \method.}
    \label{algo:compre_inf_process}     
\end{algorithm}

\section{Experiments}

In this section, we utilize \method{} to defend the state-of-the-art graph backdoor attack method to demonstrate that \method{} effectively defend against backdoor attacks while maintaining relatively high prediction accuracy.

\subsection{Experiment Setup} \label{sec:exp_setup} 

\textbf{Datasets.} For graph classification, we conducted experiments on four real-world datasets: AIDS \cite{riesen2008iam}, PROTEINS \cite{borgwardt2005protein}, PROTEINS\_full \cite{borgwardt2005protein}, and ENZYMES \cite{borgwardt2005protein}. 
For node classification, we evaluate the effectiveness of \method{} on four real-world datasets: Cora \cite{mccallum2000automating}, PubMed \cite{sen2008collective}, OGBN-Arxiv \cite{hu2020open} and Flickr \cite{zeng2019graphsaint}. The detailed description of these datasets are provided in Appendix \ref{sec:dataset_desc_appx}. 


\textbf{Baselines.} Due to limited prior work on backdoor defense for graphs, we compare \method{} with state-of-the-art backdoor defense methods in the image domain and some typical graph defense methods. Backdoor defense methods on images include ABL \cite{li2021anti}, ANP~\cite{wu2021adversarial} and fine-tune \cite{sha2022fine}. Typical graph defense methods include edge dropout and GCNJaccard \cite{wu2020graph}. Details are in Appendix \ref{sec:baseline_desc_appx}.


\textbf{Parameter Settings.} For graph classification, we employed GTA \cite{xi2021graph} as the backdoor attack method with GCN (Graph Convoluation Network) being the surrogate GNN. For node classification, we employed UGBA \cite{dai2023unnoticeable} instead of GTA \cite{xi2021graph}, since UGBA exhibits a stronger attack on node classification. The attack settings followed the default configurations, with a $10\%$ injection rate. For \method{}, we set $\lambda$ to 5, and adopted a $2$-layer GCN  as the model architecture, with a hidden dimension of 128 and $\beta$ of 0.9. The Adam optimizer was used with a weight decay of 5e-4 and betas set to 0.5 and 0.999. We set the learning rate as 0.01 and decayed by 0.1 every 40 epochs, with 200 training epochs \textit{}in total. Each experiment setting is repeated for 5 times under different random seeds, and the mean results are recorded.

\textbf{Evaluation Metrics.} The accuracy (ACC) is to assess the classification performance. A higher accuracy indicates better utility of models. The attack success rate (ASR) is to evaluate the defense effectiveness. A lower ASR indicates a smaller impact of the backdoor attack, thereby signifying a more successful backdoor defense.

\subsection{Experiment Results}

\subsubsection{Comprehensive Experimental Results} 

\begin{table*}[t]
    \centering
    \caption{Main results on graph classification. Higher is better ($\uparrow$) for ACC (white). Lower is better ($\downarrow$) for ASR (gray). Among all the defense methods, bold font indicates the best average performance for utility and backdoor defense.}
    \begin{tabular}{c c|cc | cc | cc | cc}
        \toprule
         \multirow{2}{*}{\textbf{Methods}} & \multirow{2}{*}{\textbf{Models}} & \multicolumn{2}{c}{\textbf{AIDS}} & \multicolumn{2}{c}{\textbf{PROTEINS}} & \multicolumn{2}{c}{\textbf{PROTEINS\_full}} & \multicolumn{2}{c}{\textbf{ENZYMES}} \\
         & & ASR(\%) $\downarrow$ & ACC(\%) $\uparrow$  & ASR(\%) $\downarrow$ & ACC(\%) $\uparrow$  & ASR(\%) $\downarrow$ & ACC(\%) $\uparrow$  & ASR(\%) $\downarrow$ & ACC(\%) $\uparrow$ \\
        \midrule
         \multirow{4}{*}{Vanilla} & GCN & 100.00 & 96.20 & 100.00 & 66.42 & 100.00 & 71.30 & 100.00 & 35.83 \\
         & GAT & 99.80 & 97.60 & 100.00 & 72.26 & 97.73 & 74.43 & 100.00 & 33.33 \\
         & GraphSAGE & 100.00 & 95.40 & 100.00 & 66.42 & 97.28 & 69.05 & 100.00 & 37.50 \\
         \rowcolor{lightgray!45} \cellcolor{white} & Average & 99.93 & \textbf{96.40} & 100.00 & 68.37 & 98.34 & \textbf{71.59} & 100.00 & 35.55 \\
         
        \midrule
         \multirow{4}{*}{Edge Dropout} & GCN & 95.00 & 91.60 & 100.00 & 66.42 & 100.00 & 65.92 & 98.33 & 35.00 \\
         & GAT & 99.44 & 96.12 & 100.00 & 71.97 & 97.64 & 68.70 & 100.00 & 27.17 \\
         & GraphSAGE & 99.84 & 94.88 &  100.00 & 68.02 & 97.55 & 69.41 & 100.00 & 26.67 \\
         \rowcolor{lightgray!45} \cellcolor{white} & Average & 98.09 & 94.20 & 100.00 & 68.80 & 98.40 & 68.01 & 99.44 & 29.61 \\

        \midrule
         \multirow{4}{*}{Finetune} & GCN & 92.76 & 95.36 & 100.0 & 71.53 & 100.00 & 59.64 & 39.00 & 33.67 \\
         & GAT & 89.24 & 93.36 & 74.41 & 76.64 & 100.00 & 59.64 & 37.50 & 28.33 \\
         & GraphSAGE & 89.08 & 95.04 & 100.0 & 71.53 & 43.52 & 60.00 & 19.70 & 23.50 \\
         \rowcolor{lightgray!45} \cellcolor{white} & Average & 90.36 & 94.59 & 91.47 & \textbf{73.23} & 81.17 & 59.76 & 32.07 & 28.50\\

        \midrule
         \multirow{4}{*}{ANP} & GCN & 99.80 &  94.73 & 100.00 & 69.82 & 100.00 & 75.56 & 100.00 & 38.61 \\
         & GAT & 99.20 & 98.00 & 100.00 & 72.01 & 98.79 & 70.25 & 96.11 & 37.78 \\
         & GraphSAGE & 99.80 & 96.67 & 100.00 & 70.07 & 99.40 & 75.03 & 98.89 & 50.83 \\
         \rowcolor{lightgray!45} \cellcolor{white} & Average & 99.60 & 96.47 & 100.00 & 70.63 & 99.40 & 73.61 & 98.33 & \textbf{42.41} \\

        \midrule
         \multirow{4}{*}{ABL} & GCN & 98.80 & 96.80 & 99.90 & 65.42 & 100.00 & 59.64 & 40.00 & 16.67 \\
         & GAT & 0.21 & 86.70 & 19.30 & 40.14 & 20.45 & 48.07 & 0.00 & 16.67 \\
         & GraphSAGE & 0.00 & 86.80 & 20.00 & 40.14 & 0.91 & 44.30 & 40.00 & 16.67 \\
         \rowcolor{lightgray!45} \cellcolor{white} & Average & 33.00 & 90.10 & 46.40 & 48.57 & 40.45 & 50.67 & 26.67 & 16.67\\

        \midrule
         \multirow{2}{*}{GCNJaccard} & GCN &0.00 & 86.80 & 0.00 & 33.57 & 100.00 & 59.64 & 0.00 & 21.67 \\
         & \cellcolor{lightgray!45} Average & \cellcolor{lightgray!45} 0.00 & \cellcolor{lightgray!45} 86.80 & \cellcolor{lightgray!45} 0.00 & \cellcolor{lightgray!45} 33.57 & \cellcolor{lightgray!45} 100.00 & \cellcolor{lightgray!45} 59.64 & \cellcolor{lightgray!45} 0.00 & \cellcolor{lightgray!45} 21.67 \\

        \midrule
         \multirow{4}{*}{\method{}} & GCN & 0.00 & 94.20 & 0.00 & 65.69 & 0.00 & 71.75 & 0.00 & 38.83 \\
         & GAT & 7.00 & 93.40 & 0.00 & 75.91 & 0.00 & 64.57 & 0.00 & 39.17 \\
         & GraphSAGE & 1.20 & 94.20 & 0.00 & 73.80 & 0.00 & 72.24 & 0.00 & 40.83 \\ 
         \rowcolor{lightgray!45} \cellcolor{white} & Average & \textbf{2.73} & 93.93 & \textbf{0.00} & 71.80 & \textbf{0.00} & 69.52 & \textbf{0.00} & 39.61 \\
        \bottomrule
         
    \end{tabular}
    \label{tab:compreh_experiment_result_graph_cls}
\end{table*}

\begin{table*}[t]
    \centering
    \caption{Main results on node classification. Higher is better ($\uparrow$) for ACC (white). Lower is better ($\downarrow$) for ASR (gray). Among all the defense methods, bold font indicates the best average performance for utility and backdoor defense.}
    \begin{tabular}{cc|cc | cc| cc| cc}
        \toprule
         \multirow{2}{*}{\textbf{Methods}} & \multirow{2}{*}{\textbf{Models}} & \multicolumn{2}{c}{\textbf{Cora}} & \multicolumn{2}{c}{\textbf{PubMed}} & \multicolumn{2}{c}{\textbf{OGBN-Arxiv}} & \multicolumn{2}{c}{\textbf{Flickr}} \\
         & & ASR(\%) $\downarrow$ & ACC(\%) $\uparrow$  & ASR(\%) $\downarrow$ & ACC(\%) $\uparrow$  & ASR(\%) $\downarrow$ & ACC(\%) $\uparrow$  & ASR(\%) $\downarrow$ & ACC(\%) $\uparrow$ \\
        \midrule
         \multirow{4}{*}{\textbf{Vanilla}} & GCN & 99.82 & 81.48 & 86.10 & 81.86 & 99.54 & 59.17 & 99.31 & 42.40 \\
         & GAT & 73.95 & 80.51 & 80.38 & 79.23 & 0.01 & 65.92 & 60.73 & 44.33 \\
         & GraphSAGE & 97.24 & 79.33 & 81.40 & 83.95 & 94.76 & 61.33 & 95.87 & 45.25\\
         \rowcolor{lightgray!45}
         \cellcolor{white} & Average & 90.34 & 80.44 & 82.63 & 81.68 & 64.77 & 62.14 & 85.30 & 43.99 \\
        \midrule
         \multirow{4}{*}{\textbf{Edge Dropout}} & GCN & 51.82 & 76.14 & 48.81 & 82.25 & 52.70 & 56.40 & 42.65 & 41.05 \\
         & GAT & 44.44 & 74.44 & 37.67 & 75.69 & 3.50 & 58.30 & 35.79 & 0.38 \\
         & GraphSAGE & 50.93 & 73.25 & 39.06 & 82.91 & 4.49 & 58.30 & 35.07 & 43.12 \\
         \rowcolor{lightgray!45}
         \cellcolor{white} & Average & 49.06 & 74.61 & 41.85 & 80.28 & 20.23 & 57.67 & 37.84 & 28.18\\
        \midrule
         \multirow{4}{*}{\textbf{Finetune}} & GCN & 99.46 & 83.18 & 92.66 & 81.70 & 58.57 & 48.49 & 20.00 & 32.79 \\
         & GAT & 32.35 & 81.48 & 30.70 & 69.98 & 0.02 & 48.31 & 59.52 & 26.10 \\
         & GraphSAGE & 99.28 & 80.59 & 83.16 & 84.54 & 92.94 & 51.09 & 33.03 & 27.57 \\
         \rowcolor{lightgray!45}
         \cellcolor{white} & Average & 77.03 & 81.75 & 68.84 & 78.74 & 50.51 & 49.30 & 37.52 & 28.82\\
        \midrule
         \multirow{4}{*}{\textbf{ANP}} & GCN & 51.91 & 75.86 & 74.81 & 79.99 & 50.10 & 65.72 & 35.59 & 48.37  \\
         & GAT & 55.25 & 81.09 & 73.76 & 80.49 & 48.41 & 69.65 & 24.38 & 48.90\\
         & GraphSAGE & 49.78 & 80.12 & 80.20 & 83.44 & 42.65 & 64.91 & 21.48 & 48.76 \\
         \rowcolor{lightgray!45}
         \cellcolor{white} & Average & 52.31 & 79.02 & 76.26 & 81.31 & 47.05 & 66.76 & 27.15 & 48.68 \\
        \midrule
         \multirow{4}{*}{\textbf{ABL}} & GCN & 96.88 & 80.81 & 88.24 & 78.09 & 0.00 & 35.78 & 20.00 & 31.28\\
         & GAT  & 15.11 & 73.85 & 7.50 & 60.73 & 0.00 & 35.84 & 0.02 & 26.31 \\
         & GraphSAGE & 91.02 & 68.74 & 87.73 & 72.60 & 37.92 & 34.89 & 49.33 & 27.29 \\
         \rowcolor{lightgray!45}
         \cellcolor{white} & Average & 67.67 & 74.47 & 61.16 & 70.47 & 12.64 & 35.50 & 23.12 & 28.29\\
        \midrule
         \multirow{2}{*}{\textbf{GCNJaccard}} & GCN & 99.55 & 80.74 & 12.48 & 77.13 & 7.89 & 18.87 &  2.39 & 40.51 \\
          & \cellcolor{lightgray!45} Average & \cellcolor{lightgray!45} 99.55 & \cellcolor{lightgray!45} 80.74 & \cellcolor{lightgray!45} 12.48 & \cellcolor{lightgray!45} 77.13 & \cellcolor{lightgray!45} 7.89 & \cellcolor{lightgray!45} 18.87 &  \cellcolor{lightgray!45} 2.39 & \cellcolor{lightgray!45} 40.51 \\
        \midrule
         \multirow{4}{*}{\textbf{\method{}}} & GCN & 0.00 & 83.70 & 5.71 & 71.67 & 0.00 & 54.77 & 0.00 & 41.86 \\
         & GAT & 3.73 &  80.66 & 3.38 & 81.63 & 0.06 & 63.21 & 0.82 & 36.04 \\
         & GraphSAGE & 1.33 & 64.93 & 0.84 & 85.09 & 0.00 & 61.02 & 0.00 & 45.16\\
         \rowcolor{lightgray!45}
         \cellcolor{white} & Average & 1.69 & 76.43 & 3.31 & 79.46 & 0.02 & 59.67 & 0.27 & 41.02 \\
        \bottomrule
    \end{tabular}
    \label{tab:compreh_experiment_result_node_cls}
\end{table*}

\par\noindent

We conduct comprehensive experiments on four real-world graph (node) classification datasets, and the results are detailed in Table \ref{tab:compreh_experiment_result_graph_cls} (Table \ref{tab:compreh_experiment_result_node_cls}). For both these two tasks, our proposed method exhibits the lowest Adversarial Success Rate (ASR) while maintaining satisfactory natural accuracy across all datasets. This suggests its effectiveness in defending against backdoor attacks, especially when compared to typical backdoor defense methods like ABL and ANP, which struggle to perform well on graphs. Moreover, the natural accuracy drop of our \method, when compared with natural training (vanilla), is minimal—less than $5\%$ in most cases and even better in certain scenarios, such as PROTEINS and ENZYMES. This improved natural accuracy may be attributed to the masks, which encourage original GNNs to focus more on key parts of graphs. 

Beyond the commonly used GCNs, we extended our experiments to Graph Attention Networks (GAT) \cite{velivckovic2017graph} and GraphSage \cite{hamilton2017inductive}, as detailed in Tables \ref{tab:compreh_experiment_result_graph_cls} and \ref{tab:compreh_experiment_result_node_cls}. For both tasks, our \method~ consistently defends against backdoor attack across different graph neural networks. These findings affirm the versatility of our \method{} as a universal approach for graph backdoor defense. 


\subsubsection{Defense against Backdoor Attack with Various Strengths}

\begin{figure*}[h]
    \centering
    \includegraphics[width=\textwidth]{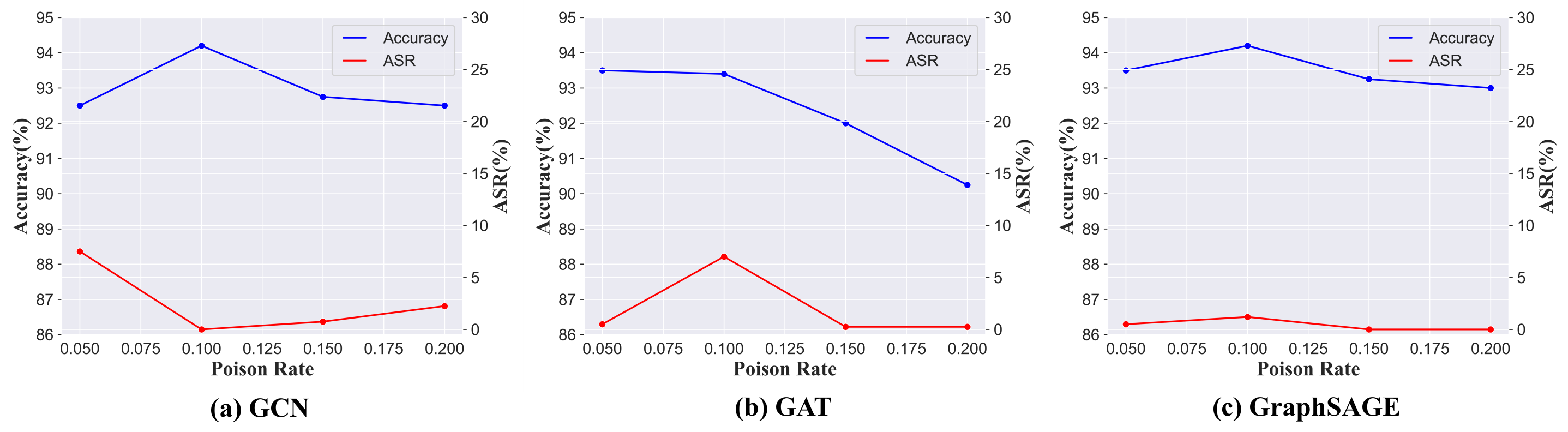}
    \caption{The impact of injection rate on Accuracy and ASR on the AIDS dataset when using different model architectures.}
    \label{fig:ablation_study}
\end{figure*}

\begin{figure*}[htbp]
    \centering
    \resizebox{\textwidth}{!}{
    \includegraphics{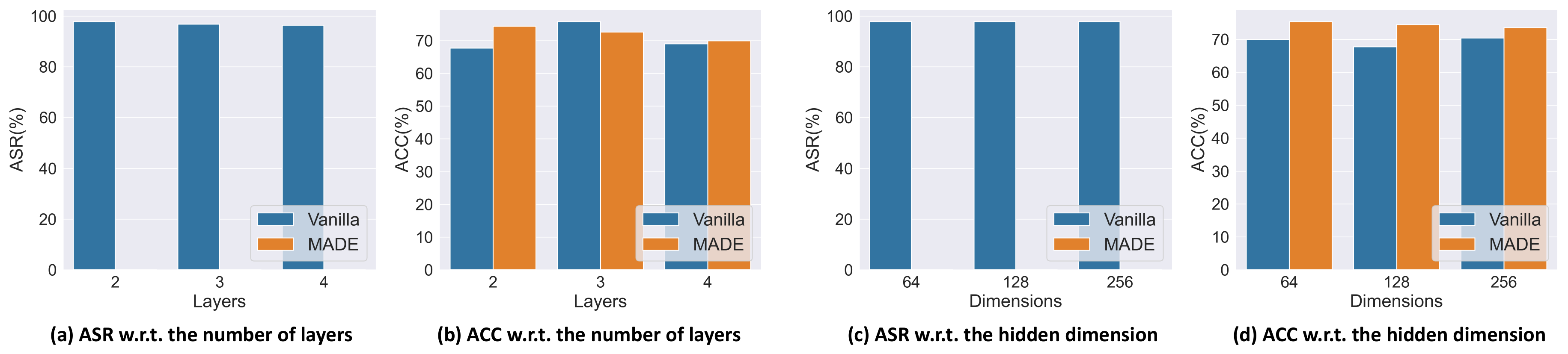}
    }
    \caption{Ablations study between vanilla training and \method{} on the PROTEINS\_full dataset for graph classification when varying the number of layers and the hidden dimension, respectively.}
    \label{fig:abl_study_hyper}
\end{figure*}

In this section, we conduct a comprehensive assessment of our proposed defense method's efficacy in countering backdoor attacks of varying strengths. Specifically, we deploy \method{} on AIDS dataset against GTA backdoor attacks with injection rates of 0.05,0.10,0.15, and 0.20. We gauge the effectiveness of our approach across GCN, GAT, and GraphSAGE, all configured with default parameters mentioned in Section \ref{sec:exp_setup}. The experiment results are shown in Figure \ref{fig:ablation_study}. Notably, we observe successful removal of backdoor triggers across all injection rates, with all the attack success rates less than 0.08 and most of them less than 0.02. Impressively, even when confronted with the backdoor attack with an injection rate of 0.20, our method maintains a competitive natural accuracy, with less than $3\%$ decrease in accuracy. This remarkable performance underscores the superiority and effectiveness of \method.



\subsubsection{Ablation Study} \label{sec:ablation_study}


To investigate the sensitivity of \method{} to hyperparameters of GNNs, we conduct a comparative analysis with GCN on both graph classification and node classification, varying the number of layers and hidden dimensions. The detailed results of node classification are provided in Appendix \ref{appdix:ablation_study_node}. Here, we mainly focus on graph classification, whose results are shown in Figure \ref{fig:abl_study_hyper}.
Specifically, to investigate the impact of the number of layers on backdoor defense, experiments are performed using 2-layer GNNs with a hidden dimension of 128, 3-layer GNNs with hidden dimensions of 128 and 32, and 4-layer GNNs with hidden dimensions of 128, 32, and 32. The corresponding results are illustrated in Figure \ref{fig:abl_study_hyper}(a) and \ref{fig:abl_study_hyper}(b). Regarding the model's hidden layer dimensions, given a 2-layer GNN, we conducted extensive experiments with hidden dimensions set to 64, 128, and 256, whose results are depicted in Figure \ref{fig:abl_study_hyper}(c) and \ref{fig:abl_study_hyper}(d). The results show that \method{} reduces the attack success rate to near-zero while maintaining high accuracy across all the configurations, demonstrating the insensitivity of \method~ on hyperparameters. Notably, under specific hyperparameter configurations, such as 2 or 4 layers or a hidden dimension of 64, \method{} consistently demonstrates improved performance in both backdoor defense and utility, highlighting its superiority.

\subsubsection{Evaluations on Unlearning with Masks}

\begin{figure}
    \centering
    \begin{subfigure}{0.48\linewidth}
        \resizebox{\linewidth}{!}{
            \includegraphics{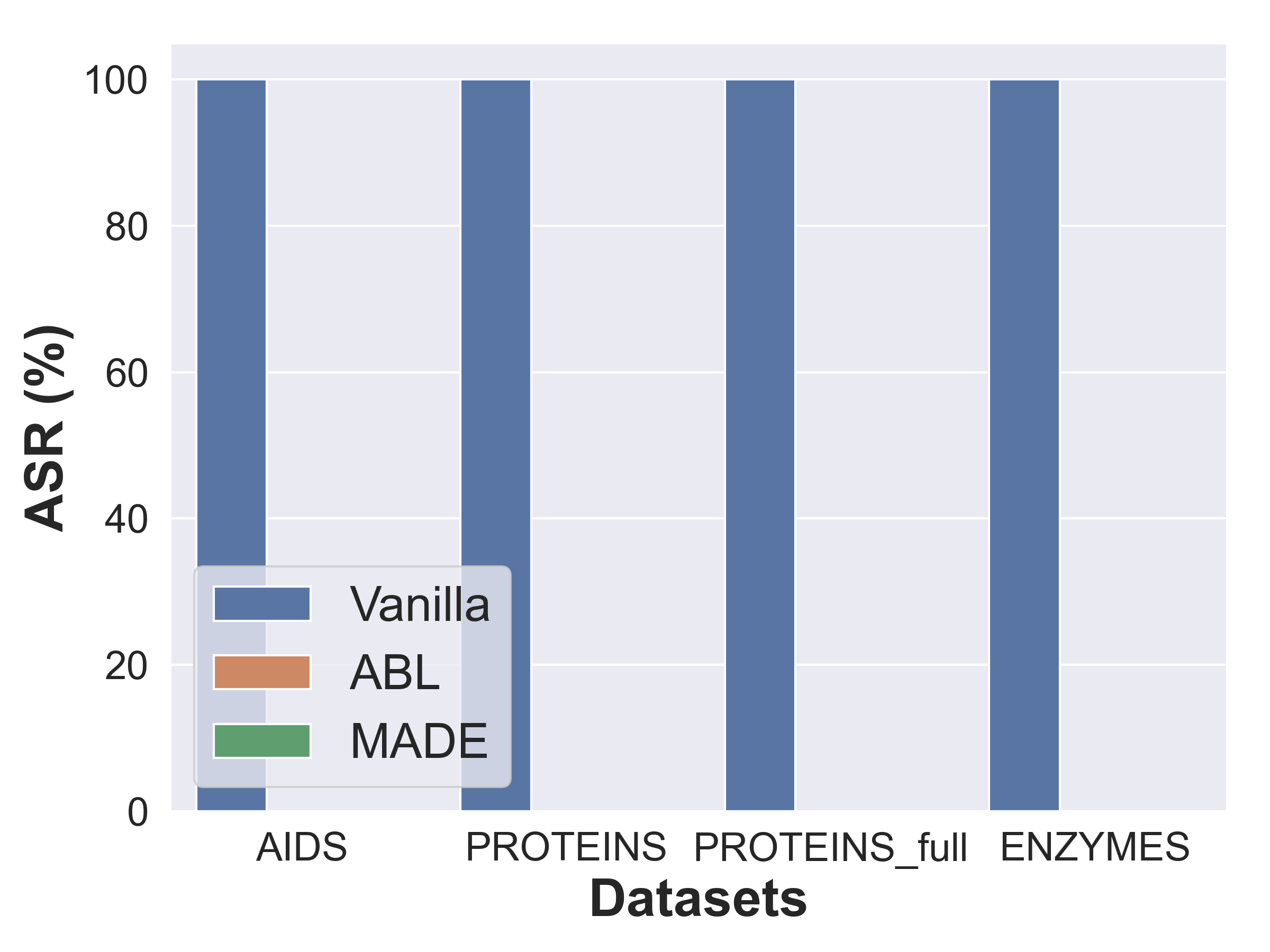}
        }
        \caption{ASR w.r.t. datasets}
    \end{subfigure}
    \begin{subfigure}{0.48\linewidth}
        \resizebox{\linewidth}{!}{
            \includegraphics{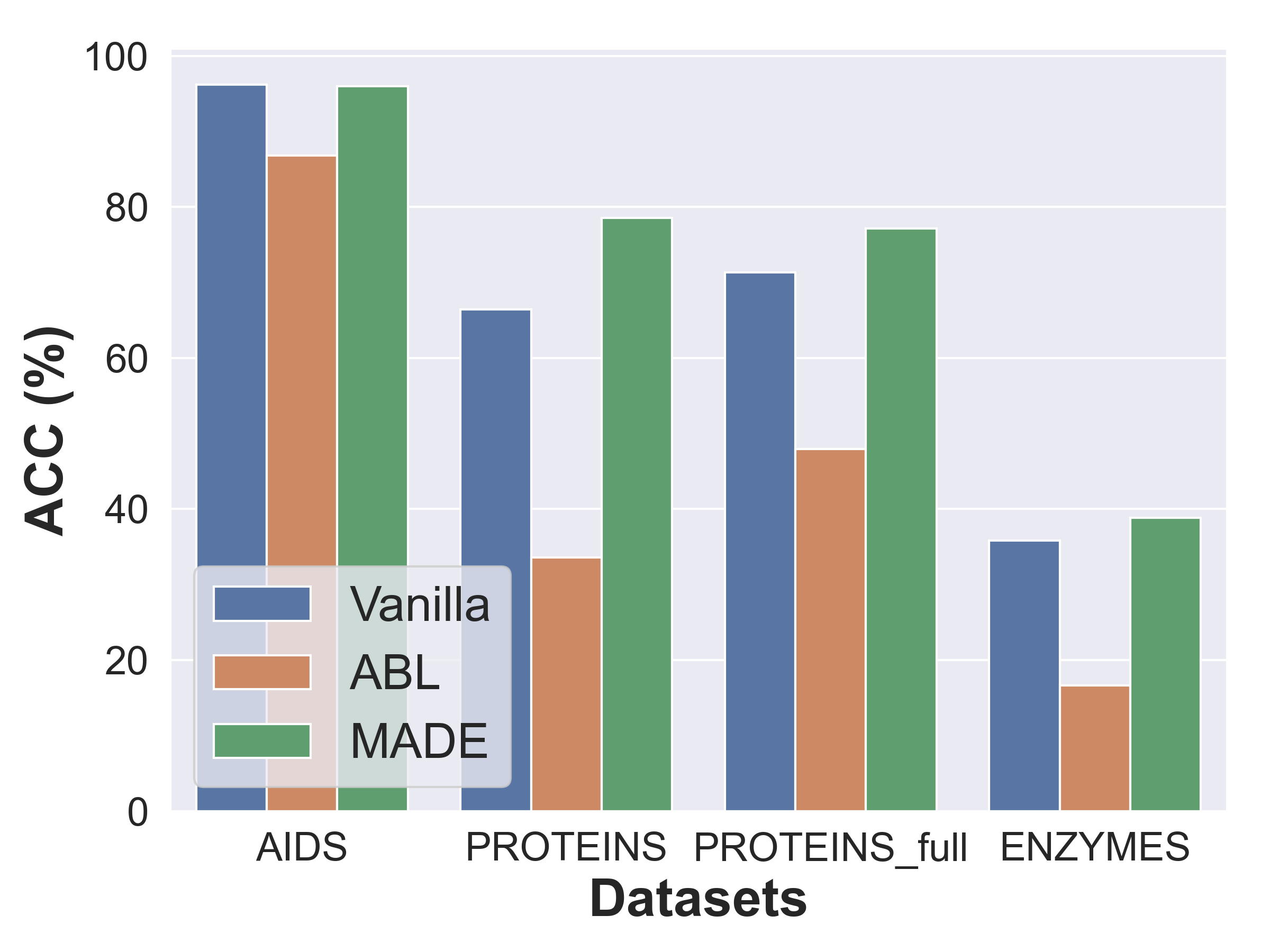}
        }
        \caption{ACC w.r.t. datasets}
    \end{subfigure}
    \caption{Ablation study to show the effectiveness of mask unlearning. We utilize three methods (Vanilla training, ABL, and \method{}) to train a GCN on four graph classification datasets purified with the same data isolation. Higher is better ($\uparrow$) for ACC. Lower is better ($\downarrow$) for ASR. }
    \label{fig:maskcomp_data_iso}
\end{figure}

In this subsection, we aim to further validate the conclusion in this paper that a well-designed mask generation module can help overcome the over-modification problem of original unlearning, thus ensuring better performance. Therefore, we compare the performance of ABL and \method{} when trained on the same purified training samples which are obtained through data isolation in this paper.

From Figure \ref{fig:maskcomp_data_iso}, it is evident that ABL can still eliminate the influence of backdoor attacks when employing better data isolation methods. However, this comes at the expense of a significant decrease in accuracy, supporting the conclusion that original unlearning may excessively modify neurons in networks.
In contrast, \method{} consistently achieves near-zero ASR while maintaining high classification accuracy across all four datasets. Notably, under some datasets (PROTEINS\_full and ENZYMES for data isolation, and PROTEINS, PROTEINS\_full and ENZYMES for ground truth), the accuracy of \method{} even surpasses that of the vanilla GCN. It clearly demonstrates the superiority of \method{}. This superior performance may be attributed to the fact that, with edge masking, only the poisoned nodes in an attacked graph are eliminated, preserving the topological structure between the remaining clean nodes.

\section{Related Works}
\subsection{Graph Neural Networks.} 
Graph neural network (GNN) has demonstrated state-of-the-art performance in various learning tasks, such as drug discovery \citep{lim2019predicting, lin2020kgnn, jiang2021could}, traffic forecasting \citep{jiang2022graph}, 3D object detection \citep{shi2020point}, recommender systems \citep{fan2019graph, wang2019neural}, and webpage ranking \citep{bojchevski2019pagerank, klicpera2018predict}. Graph Convolutional Network (GCN) \citep{kipf2016semi} utilizes localized first-order spectral graph convolution to learn node representations. GraphSAGE\citep{hamilton2017inductive} uses neighborhood sampling and aggregation to employ graph neural networks in inductive learning. GAT\citep{velivckovic2017graph} leverages self-attention mechanisms to learn attention weights for neighborhood aggregation. 

\subsection{Backdoor Attack and Defenses}
\par\noindent\textbf{Backdoor Attacks.}
Existing backdoor attacks aim to optimize three objectives: 1) making the trigger more invisible, 2) reducing the trigger injection rate, and 3) increasing the attack success rate. Many works focus on designing special patterns of triggers to make them more invisible. Triggers can be designed as simple patterns, such as a single pixel \citep{tran2018spectral} or a black-and-white checkerboard \citep{gu2017badnets}. Triggers can also be more complex patterns, such as mixed backgrounds \citep{chen2017targeted}, natural reflections \citep{liu2020reflection}, invisible noise \citep{liao2018backdoor}, and adversarial patterns \citep{zhao2020clean}. 
As for backdoor attacks on graphs, GTA \cite{xi2021graph} dynamically adapts triggers to individual graphs by optimizing both attack effectiveness and evasiveness on a downstream classifier. TRAP \cite{yang2022transferable} generates perturbation triggers via a gradient-based score matrix from a surrogate model. UGBA \cite{dai2023unnoticeable} deploys a trigger generator to inject unnoticeable triggers into the nodes deliberately selected for stealthiness and effectiveness.

\par\noindent\textbf{Backdoor Defense.} 
In the field of backdoor defense, \cite{wang2019neural, wang2022rethinking} tries to reverse potential triggers through imitation and reverse engineering of backdoor attacks, then they try to remove the generated triggers to enhance defense effectiveness. Apart from these reverse-based methods, \cite{wu2021adversarial, guan2022few} try to identify malicious neurons and purify backdoored models by exploiting the sensitivity of backdoored neurons to adversarial perturbations. Besides the above methods, \cite{li2021anti, huang2022backdoor,tran2018spectral} try to detect backdoor samples and then remove the poison behavior by retraining or unlearning. However, those existing methods are not suitable for the graph scenario. Apart from these image-based methods, \cite{jiang2022defending} is designed for graph datasets. However, such a method only focuses on detecting the injected edge trigger and can not deal with feature triggers or node triggers. From the above illustration, one can see the effective backdoor defense method for graph neural networks are urgently needed.



\section{Conclusion}
In this paper, we research on the usefulness of graph topology on backdoor defenses. We empirically demonstrate the drawback of traditional backdoor defense methods on graphs, and further give insightful explanations. Guided by our analyses, we point out that mask learning serves as an excellent graph defense framework to remain the clean part of graphs while mitigating malicious effect of triggers. Hence, we propose \method{}, which utilizes graph topology to generate defensive masks for the purpose of preserving untainted subgraphs and hence training trigger-insensitive robust GNNs. Extensive evaluations on real-world datasets have demonstrated the effectiveness of \method{} on defending backdoor attack while maintaining competitive utility.

\bibliographystyle{IEEEtran}
\bibliography{ref}

\begin{IEEEbiography}[{\includegraphics[width=1in,height=1.25in,clip,keepaspectratio]{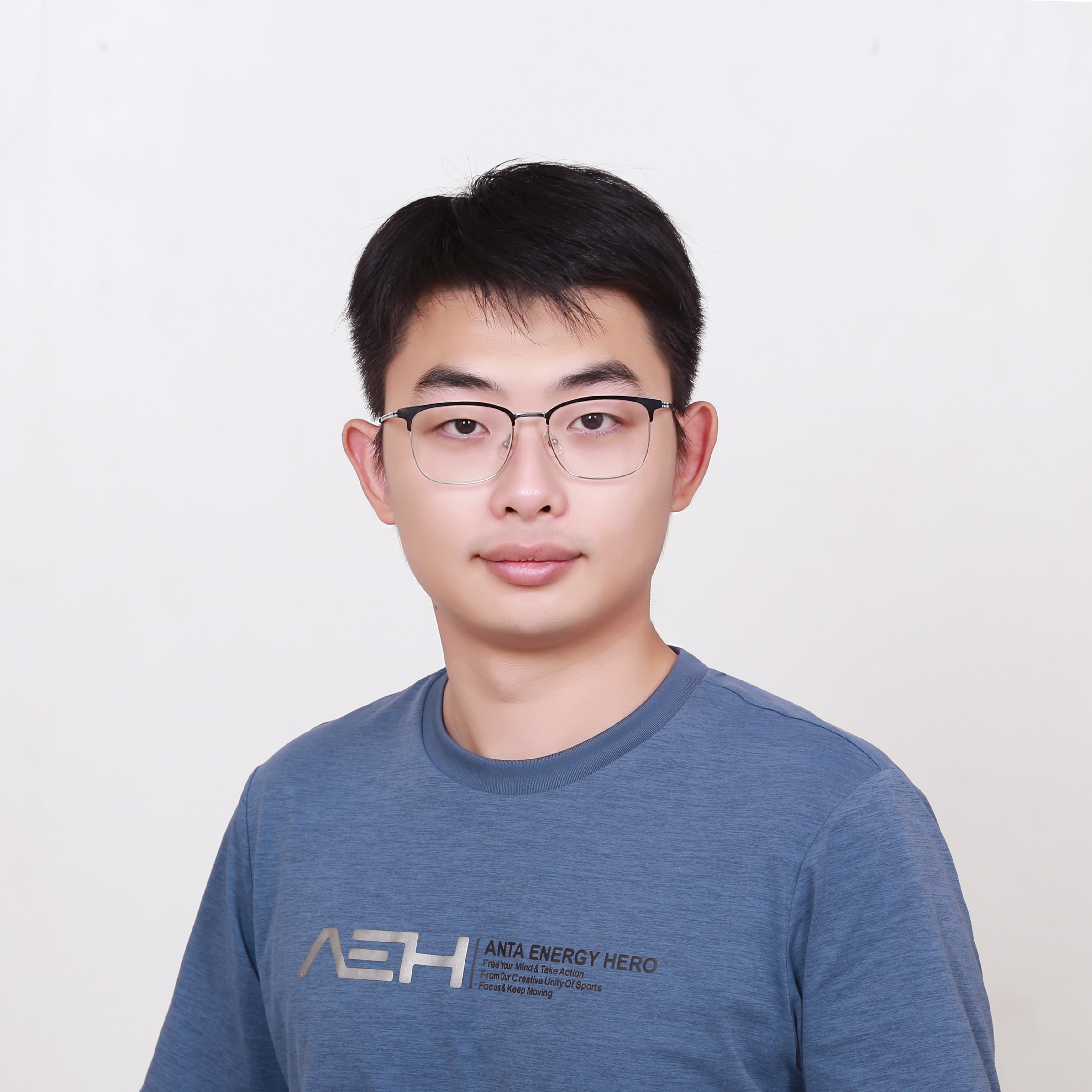}}]{Xiao Lin} received the B.E. degree from Peking University in 2019. He is currently a Ph.D. student in IdeaLab, University of Illinois Urbana-Champaign. His research interests include graph learning, trustworthy machine learning, and time series analysis.
\end{IEEEbiography}

\begin{IEEEbiography}[{\includegraphics[width=1in,height=1.25in,clip,keepaspectratio]{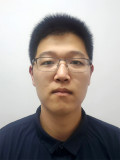}}]{Mingjie Li}
received the Ph.D. degree from Peking University in 2023. His research interest includes trustworthy machine learning, such as adversarial robustness, privacy, and data security.
\end{IEEEbiography}


\begin{IEEEbiography}
[{\includegraphics[width=1in,height=1.25in,clip,keepaspectratio]{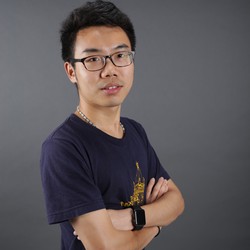}}]{Yisen Wang} received the Ph.D. degree from Tsinghua University in 2018. He is currently an Assistant Professor at Peking University. His research interest includes machine learning and deep learning, such as adversarial learning, graph learning, and weakly/self-supervised learning.
\end{IEEEbiography}
\newpage
\appendices

\section{\method{} on Node Classification}
\label{appdx:node-class-made}

The \method{} algorithm for node classification is highly similar to that for graph classification. However, due to the differences between these tasks, we have made slight modifications to \method{} to enhance its defensive efficacy. These modifications primarily focus on (1) data isolation and (2) masked aggregation.

\textbf{Data Isolation}. To achieve node-level data isolation, we need to adapt the graph-level homophily in Eq. \eqref{eq:graph_homo} to a node-specific homophily metric. First, we leverage ego graphs to determine node homophily. For each node $v$, we extract its 2-hop ego graph $\mathcal{G}_v$ and utilize the homophily of the ego graph as the node homophily. Second, we aim to use the uniqueness of node classification to improve data isolation. Specifically, for node classification, since backdoor attacks significantly alter the predictions for attacked nodes, the pseudo-labels of attacked nodes may markedly differ from those of neighboring nodes. Consequently, calculating homophily based on the similarity of pseudo-labels can yield more effective data isolation. Hence, for the node $v$, Eq. \eqref{eq:graph_homo} could be modified as:
\begin{equation} \label{eq:node_homo}
    \textrm{homo}(v) = \frac{1}{\vert \mathcal{N}_v \vert} \sum_{j \in \mathcal{N}_v} \textrm{sim}(\hat{y}_v, \hat{y}_j)
\end{equation}
where $\mathcal{N}_v$ is the neighboring nodes within the ego graph $\mathcal{G}_v$, $\hat{y}_j$ is the pseudo-label of the node $j$. Once the node homophily is clearly defined, we are able to calculate $\homoMean$ and $\homoStd$, and then implement data isolation strictly following the rest part of Section \ref{sec:data_isolation}.
 
\textbf{Masked Aggregation}. After detecting the trigger nodes using our data isolation method, we then try to remove the impacts of these poisoned triggers. To achieve such a goal, we design masks on the graph adjacency matrix to remove the trigger nodes or edges while maintaining the original graph structure. The masks are defined as follows:

\begin{equation} \label{eq:mask_express_node}
    \mathbf{m}^{(k)}_{i, j} = \left\{
    \begin{aligned}
        A_{ij},& \qquad v_i \in \clnV \  and \  v_j \in \clnV \\
        0,& \qquad else,
    \end{aligned}
    \right.
\end{equation}
where $\mathcal{V}_{clean}$ denoted the vertex set containing all the benign nodes split in our data isolation part.
After obtaining $\mathbf{m}$, we treat the graph as a weighted graph for message passing. Mathematically, the message passing is modified as follows:
\begin{align}
    &\mathbf{a}_i^{(k)} = \textbf{AGGREGATE} \left(\{\mathbf{m}_{i, j}^{(k)} \cdot \mathbf{h}_j^{(k)}: v_j \in \mathcal{N}(v_i)\} \right) \label{mask-agg}\\
    &\mathbf{h}_i^{(k+1)} = \textbf{COMBINE} \left(\{\mathbf{m}_{i, i}^{(k)} \label{mask-comb}\cdot \mathbf{h}_i^{(k)}, \mathbf{a}_i^{(k)}\} \right)
\end{align}

Once the masked aggregation is defined, we will update the model for node classification based on the  losses in Section \ref{sec:loss_define}.

\section{Evaluations on data isolation under different isolation rates.} \label{sec:data_iso_diff_rate_appx}


\begin{figure}
    \centering
    \begin{subfigure}{0.48\linewidth}
        \resizebox{\linewidth}{!}{
        \includegraphics{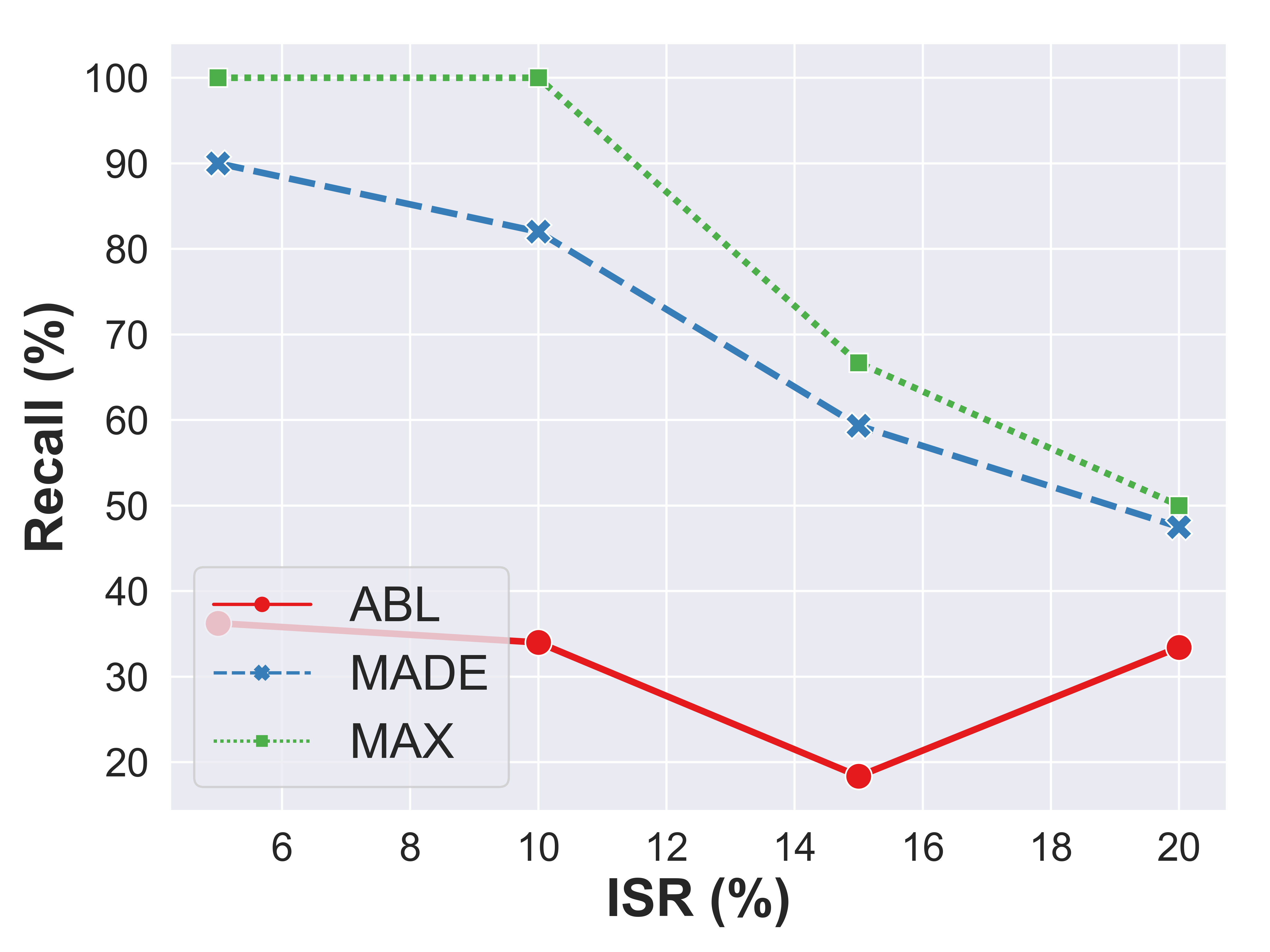}
        }
        \caption{Recall w.r.t. ISR}
    \end{subfigure}
    \begin{subfigure}{0.48\linewidth}
        \resizebox{\linewidth}{!}{
        \includegraphics{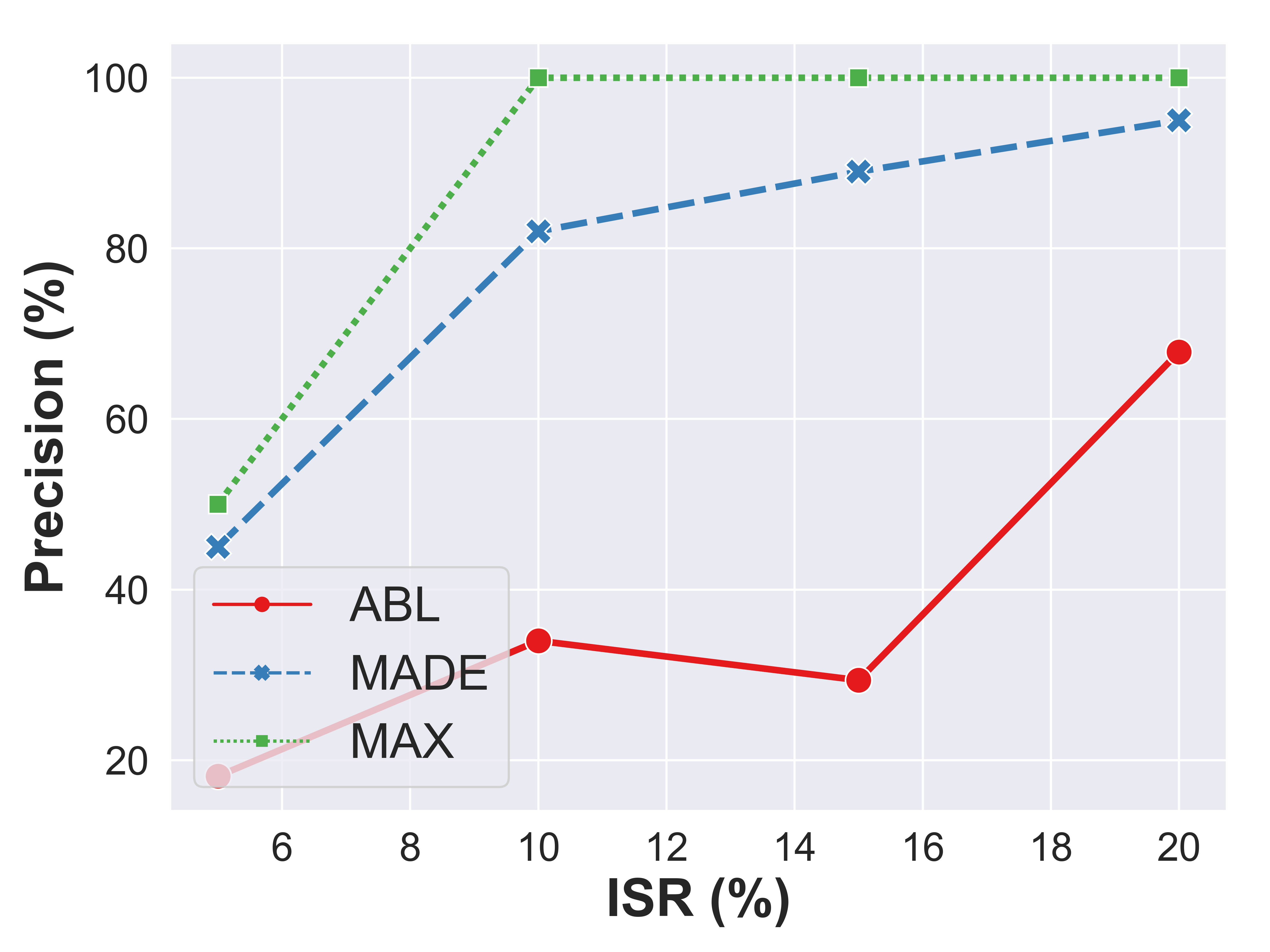}
        }
        \caption{Precision w.r.t. ISR}
    \end{subfigure}
    \caption{Precision and recall of isolated backdoor samples with changing isolation rates (ISR) on AIDS. The higher the precision and recall, the more effective the data isolation is. The maximum possible values for precision and recall are denoted as ``MAX'' (green dash line). }
    \label{fig:prec&rec_diff_rate}
\end{figure}

To validate the efficacy of data isolation, we compute the precision and recall of isolated backdoor samples among all the backdoor samples across various isolation rates. Specifically, we maintain a fixed injection rate of $10\%$ on the AIDS dataset, while adjusting the isolation rate ($5\%$, $10\%$, $15\%$, $20\%$) to compare data isolation of ABL and \method{}. The results are presented in Figure \ref{fig:prec&rec_diff_rate}. Please note that when the isolation rate changes, the maximum values for both recall and precision may vary. Mathematically, we can easily have the maximum value of precision and recall as follows:
\begin{equation}
    Precision_{max} = \max \{ 1, \frac{ISR}{IJR} \},
\end{equation}
\begin{equation}
    Recall_{max} = \max \{ 1, \frac{IJR}{ISR} \},
\end{equation}
where $ISR$ represents the isolation rate and $IJR$ represents the injection rate. For ease of comparison, we have denoted their maximum values in Figure \ref{fig:prec&rec_diff_rate} under the ``Maximum'' row. Notably, at any isolation rate, \method{} consistently outperforms ABL in both precision and recall, underscoring the effectiveness of \method{} on data isolation.


\section{Evaluations on the effectiveness of individual methods in data isolation.} \label{sec:data_iso_indiv_method_appx}

\begin{table}[h]
    \centering
    \caption{Precision and recall of individual methods used to select backdoor samples. The higher the precision and recall, the more effective the method is. }
    \resizebox{\linewidth}{!}{
    \begin{tabular}{c|cccc|cccc}
        \toprule
         \multirow{2}{*}{isolation rate } & \multicolumn{4}{c}{Recall ($\uparrow$)} & \multicolumn{4}{c}{Precision ($\uparrow$)}  \\
        \cline{2-9}
         & 5\% & 10\% & 15\% & 20\% & 5\% & 10\% & 15\% & 20\% \\
        \midrule
         Loss Detection & 37.00 & 38.00 & 39.33 & 38.00 & 18.50 & 38.00 & 59.00 & 76.00\\
         Homophily Detection & 89.00 &	81.00 & 58.67 & 45.00 & 44.50 & 81.00 & 88.00 & 90.00 \\
         Data Isolation & \textbf{90.00} & \textbf{82.00} & \textbf{59.33} & \textbf{47.50} & \textbf{45.00} & \textbf{82.00} & \textbf{89.00} & \textbf{95.00} \\
         Maximum & 100.00& 100.00& 66.67	& 50.00 & 50.00 & 100.00 & 100.00 & 100.00 \\
        \bottomrule
    \end{tabular}
    }
    \label{tab:prec&rec_indiv_appx}
\end{table}

In this subsection, our objective is to assess the effectiveness of the two components of data isolation in this paper: loss detection and homophily detection. To achieve this, we conduct experiments of precision and recall on the AIDS dataset, separately leveraging loss and homophily to select backdoor samples. The experiment results are presented in Table \ref{tab:prec&rec_indiv_appx}. Please note that, when considering homophily scores in this paper, graphs falling within the region defined by Eq. (2) are considered backdoor samples. As a consequence, the number of selected backdoor samples remains fixed, posing challenges for a comprehensive comparison with the loss detection method and overall data isolation method. To address this limitation, given the isolation rate, we choose the top graphs with the lowest  homophily scores from all graphs. The experimental results are shown in Table \ref{tab:prec&rec_indiv_appx}. The results obtained solely using the loss function are labeled as ``Loss Detection'', those obtained solely using homophily are labeled as ``Homophily Detection'', the results using them both are labeled as ``Data Isolation'' and the maximum value of precision and recall are labels as ``maximum''. The results indicate that employing only the loss function or homophily separately is effective in isolating samples. However, at any isolation rate, the overall performance is always the best when both are utilized, which demonstrate the superiority of data isolation in this paper.

\section{Ablation Study on Hyperparameters to GNNs for Node Classification} \label{appdix:ablation_study_node}

\begin{figure*}[htbp]
    \centering
    \resizebox{\textwidth}{!}{
    \includegraphics{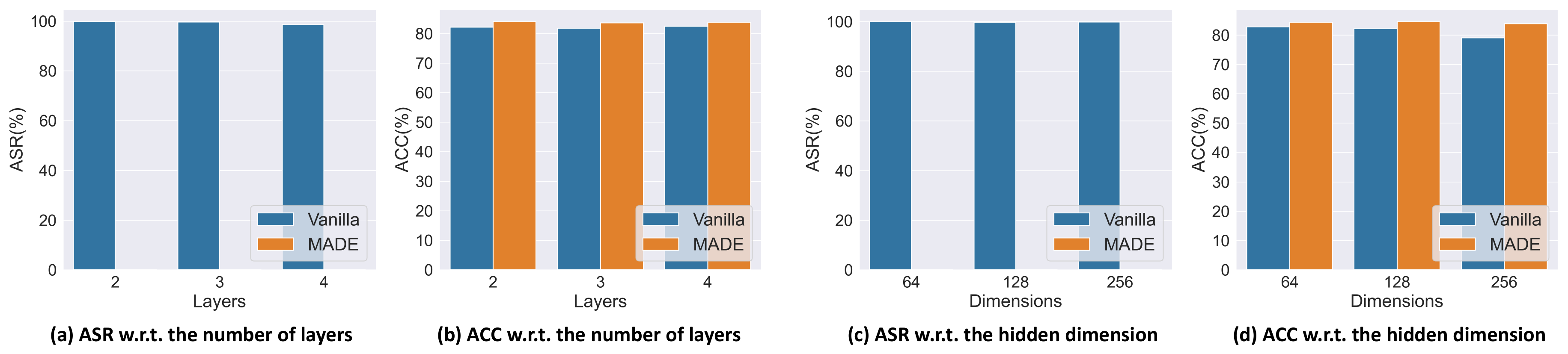}
    }
    \caption{Ablations study between vanilla training and \method{} on the Cora dataset for node classification when varying the number of layers and the hidden dimension, respectively.}
    \label{fig:abl_study_hyper_node}
\end{figure*}

We conduct comprehensive experiments to assess the hyperparameter sensitivity of \method{} in the context of node classification. The findings are illustrated in Figure \ref{fig:abl_study_hyper_node}, using the same experimental setup as described in Section \ref{sec:ablation_study} for graph classification. The results allow us to draw a parallel conclusion to that of Section \ref{sec:ablation_study}: \method{} consistently provides effective backdoor defense across neural networks with different parameter sizes, while maintaining competitive accuracies. These findings robustly demonstrate that \method{} exhibits insensitivity to hyperparameters, thereby underscoring its superior performance and reliability.

\section{Dataset Descriptions} \label{sec:dataset_desc_appx}
Here we provide a detailed descriptions for the datasets in this paper. For graph classification, we choose four widely-used dataset, \ie,  AIDS \cite{riesen2008iam}, PROTEINS \cite{riesen2008iam}, PROTEINS\_full \cite{borgwardt2005protein}, and ENZYMES \cite{borgwardt2005protein}.
\begin{itemize}
    \item AIDS \cite{riesen2008iam} comprises molecular structure graphs representing active and inactive compounds, totaling 2000 graphs. These compounds are derived from the AIDS Antiviral Screen Database of Active Compounds, which includes 4395 chemical compounds. Among these, 423 belong to class CA, 1081 to CM, and the remaining compounds to CI.
    \item PROTEINS and PROTEINS\_full \cite{borgwardt2005protein} consist of molecular structure depicting large protein molecules used for enzyme presence prediction. These datasets classify proteins as enzymes or non-enzymes, where nodes represent amino acids, and edges connect two nodes if they are less than 6 Angstroms apart. 
    \item ENZYMES \cite{borgwardt2005protein} consists of 600 protein tertiary structures sourced from the BRENDA enzyme database, featuring a total of 6 enzymes.
\end{itemize}

For node classification, we choose four real-world datasets, \ie, Cora \cite{mccallum2000automating}, PubMed \cite{sen2008collective}, OGBN-Arxiv \cite{hu2020open} and Flickr \cite{zeng2019graphsaint}.
\begin{itemize}
    \item Cora \cite{mccallum2000automating} consists of 2708 scientific publications classified into one of seven classes. The citation network consists of 5429 links. Each publication in the dataset is described by a 0/1-valued word vector indicating the absence/presence of the corresponding word from the dictionary. The dictionary consists of 1433 unique words.
    \item PubMed\cite{sen2008collective} contains 19,717 scientific publications on diabetes from the PubMed database, classified into three categories. The citation network has 44,338 links. Each publication is described by a TF/IDF weighted word vector from a dictionary of 500 unique words.
    \item The OGBN-ArXiv dataset \cite{hu2020open} is a directed graph representing the citation network of all Computer Science (CS) ArXiv papers. Nodes correspond to papers, and directed edges indicate citations. Each paper has a 128-dimensional feature vector, derived by averaging the embeddings of words in its title and abstract.
    \item Flickr \cite{zeng2019graphsaint} is derived from four sources, including NUS-WIDE, forming an undirected graph. Each node represents an image on Flickr, and edges connect images with shared properties (e.g., same location, gallery, or user comments). Node features are 500-dimensional bag-of-word representations from NUS-WIDE.
\end{itemize}

\section{Descriptions of Baseline Methods} \label{sec:baseline_desc_appx}
Here we give detailed explanations about the baselines in this paper. As for the state-of-the-art backdoor defense methods on images,

\begin{enumerate}
    \item fine-tune \cite{sha2022fine}, a widely used method in transfer learning, can effectively remove backdoors from machine learning models. In this paper, the experiment of fine-tuning runs on the clean subset isolated by data isolation.
    \item ABL \cite{li2021anti} isolate backdoor examples at an early training stage, and then break the correlation between backdoor examples and the target class by unlearning mechanism.
    \item ANP \cite{wu2021adversarial} prunes sensitive neurons to remove the injected backdoor by identifying the increased sensitivity of backdoored DNNs to adversarial neuron perturbations.
\end{enumerate}

For typical graph defense methods, 

\begin{enumerate}
    \item Edge dropout randomly remove the connections between nodes, thus decreasing the influence of poisoned nodes during message passing and defending backdoor attacks.
    \item GCN-SVD \cite{peng2022svd} replaces the core design of GCN-based methods with a low-rank truncated SVD. This modification aims to concentrate solely on the features associated with the K-largest singular vectors, thereby enhancing the model's robustness to perturbations.
\end{enumerate}

\section{\method{} Training Algorithm} 
\begin{algorithm}[h] 
    \SetKwInOut{Input}{Input}
	\SetKwInOut{Output}{Output}
    \Input{a graph dataset $\mathcal{D}$, a $K$-layer GNN with parameter $\Theta$, the epochs of warm-up phase $epoch_{warm}$, the epochs of training phase $epoch_{train}$.}
    \Output{a backdoor-insensitive GNN with parameter $\Theta_{opt}$.}
    Initialize gradient-based optimizer $optim$\;
    \tcp{warm-up phase to split datasets.}
    Calculate $\homoMean$ and $\homoStd$ based on Eqs. \eqref{eq:graph_homo} or \eqref{eq:node_homo}\;
    Form $\mathcal{D}_h$ with samples in $(0, \homoMean - \homoStd)$ $ \cup (\homoMean+\homoStd,1)$\;
    \For{epoch = $1 \rightarrow epoch_{warm}$}{
        \For{every sample in $\mathcal{D}$}{
        calculate $\natloss$ based on Eq. \eqref{eq:natural_loss}\;
        }
        update $\Theta$ by $optim$\;
    }
    Select top $\alpha_1$ samples with the lowest loss as $\mathcal{D}_h$\;
    Get $\mathcal{D}_{bad} = \mathcal{D}_{l} \cup \mathcal{D}_{h}$\;
    Select top $\alpha_2$ samples with the highest loss as $\mathcal{D}_{clean}$\;
    \tcp{train GNN with masks}
    \For{every sample in $\mathcal{D}_{clean} \cup \mathcal{D}_{bad}$}{
        \uIf{Doing graph classification}{
        Get layer $k$'s mask $\mathbf{m}^{(k)}$ based on Eq. \eqref{eq:mask_express}\;
        }
        \uElseIf{Doing node classification} {
            Get layer $k$'s mask $\mathbf{m}^{(k)}$ based on Eq. \eqref{eq:mask_express_node}\;
        }
        
        \If{$\mathcal{G} \in \mathcal{D}_{bad}$}{
            Calculate $\advloss$ based on Eq. \eqref{eq:adv_loss}\;
        }
        \Else{
            Calculate $\clnloss$ based on Eq. \eqref{eq:cln_loss}\;
        }
        update $\Theta$ by $optim$\;
    }
    $\Theta_{opt} = \Theta$
    \caption{\method~ Training Algorithm.}
    \label{algo:training}     
\end{algorithm}

\end{document}